\documentclass[traditabstract,longauth]{aa}
\usepackage{times,epsfig,amsmath}
\usepackage{graphicx}
\usepackage{epstopdf}
\usepackage{pdflscape}
\usepackage{adjustbox}
\usepackage{amssymb, amsmath} 

\usepackage[usenames,dvipsnames]{color}
\usepackage{txfonts}
\usepackage{microtype}
\usepackage{url}
\usepackage{booktabs}
\usepackage{subfig}
\usepackage{commath}
\usepackage{natbib}
\bibpunct{(}{)}{;}{a}{}{,} 
\usepackage[bookmarks, colorlinks, breaklinks, ]{hyperref}  
\hypersetup{linkcolor=blue,citecolor=MidnightBlue,filecolor=black,urlcolor=MidnightBlue} 
\newfont{\gwpfont}{cmssq8 scaled 1000}
\newcommand{\rexcess}{{\gwpfont REXCESS}}
\newcommand{\cxo}{{\it Chandra}}

\newcommand{\planck}{{\it Planck}}
\newcommand{\xmm}{{XMM-{\it Newton}}}
\newcommand{\xmms}{{\it XMM}}

\def\Mgv{M_{\rm g,500}}

\def\YX {Y_{\rm X}}
\def\YSZ {Y_{\rm SZ}}
\def\TX {T_{\rm X}}
\def\LX {L_{\rm X}}

\def\Mv {M_{\rm 500}}
\def \Rv {R_{500}}
\def\keV {\rm keV}

\def\msol {{\rm M_{\odot}}}

\begin{document}

\title{The Cluster HEritage project with XMM-\textit{Newton:}}
\subtitle{Mass Assembly and Thermodynamics at the Endpoint of structure formation \\ I. Programme overview}

\titlerunning{The Cluster HEritage project with XMM-\textit{Newton} I.}

\author{
The CHEX-MATE Collaboration:
M. Arnaud\inst{\ref{inst1}}, 
S. Ettori\inst{\ref{inst2},\ref{inst3}}, 
G.W. Pratt\inst{\ref{inst1}}, 
M. Rossetti\inst{\ref{inst4}}, 
D. Eckert\inst{\ref{inst5}}, 
F. Gastaldello\inst{\ref{inst4}}, 
R. Gavazzi\inst{\ref{inst6}}, 
S.T. Kay\inst{\ref{inst7}}, 
L. Lovisari\inst{\ref{inst2},\ref{inst8}}, 
B.J. Maughan\inst{\ref{inst9}}, 
E. Pointecouteau\inst{\ref{inst10}}, 
M. Sereno\inst{\ref{inst2},\ref{inst3}}, 
I. Bartalucci\inst{\ref{inst1},\ref{inst4}}, 
A. Bonafede\inst{\ref{inst11},\ref{inst12},\ref{inst13}},
H. Bourdin\inst{\ref{inst14}}, 
R. Cassano\inst{\ref{inst12}}, 
R.T. Duffy\inst{\ref{inst9}}, 
A. Iqbal\inst{\ref{inst1}}, 
S. Maurogordato\inst{\ref{inst15}}, 
E. Rasia\inst{\ref{inst16},\ref{inst17}},  
J. Sayers\inst{\ref{inst18}},
F. Andrade-Santos\inst{\ref{inst8}},
H. Aussel\inst{\ref{inst1}}, 
D.J. Barnes\inst{\ref{inst19}}, 
R. Barrena\inst{\ref{inst20},\ref{inst21}},  
S. Borgani\inst{\ref{inst22},\ref{inst16},\ref{inst17},\ref{inst23}}, 
S. Burkutean\inst{\ref{inst12}}, 
N. Clerc\inst{\ref{inst10}}, 
P.-S. Corasaniti\inst{\ref{inst24}, \ref{inst25}}, J.-C. Cuillandre\inst{\ref{inst1}}, 
S. De Grandi\inst{\ref{inst26}}, 
M. De Petris\inst{\ref{inst27}}, 
K. Dolag\inst{\ref{inst28}, \ref{inst29}}, 
M. Donahue\inst{\ref{inst30}}, 
A. Ferragamo\inst{\ref{inst27}}, 
M. Gaspari\inst{\ref{inst2}, \ref{inst31}}, 
S. Ghizzardi\inst{\ref{inst4}}, 
M. Gitti\inst{\ref{inst11}, \ref{inst12}}, 
C.P. Haines\inst{\ref{inst45}}, 
M. Jauzac\inst{\ref{inst32},\ref{inst33},\ref{inst34},\ref{inst35}}, 
M. Johnston-Hollitt\inst{\ref{inst36}, \ref{inst37}}, 
C. Jones\inst{\ref{inst8}}, 
F. Kéruzoré\inst{\ref{inst38}}, 
A.M.C. LeBrun\inst{\ref{inst24},\ref{inst1}}, 
F. Mayet\inst{\ref{inst38}}, 
P. Mazzotta\inst{\ref{inst14}}, 
J.-B. Melin\inst{\ref{inst39}}, 
S. Molendi\inst{\ref{inst4}}, 
M. Nonino\inst{\ref{inst16}}, 
N. Okabe\inst{\ref{inst40}}, 
S. Paltani\inst{\ref{inst5}}, 
L. Perotto\inst{\ref{inst38}}, 
S. Pires\inst{\ref{inst1}}, 
M. Radovich\inst{\ref{inst47}},
J.-A. Rubino-Martin\inst{\ref{inst20}, \ref{inst21}},
L. Salvati\inst{\ref{inst16},\ref{inst17}}, 
A. Saro\inst{\ref{inst22},\ref{inst16},\ref{inst17},\ref{inst23}}, 
B. Sartoris\inst{\ref{inst16},\ref{inst17}},
G. Schellenberger\inst{\ref{inst8}}, 
A. Streblyanska\inst{\ref{inst20}, \ref{inst21}}, 
P. Tarr\'io\inst{\ref{inst1},\ref{inst48}}, 
P. Tozzi\inst{\ref{inst41}}, 
K. Umetsu\inst{\ref{inst42}}, 
R.F.J. van der Burg\inst{\ref{inst43}, \ref{inst1}}, 
F. Vazza\inst{\ref{inst11},\ref{inst12},\ref{inst13}}, 
T. Venturi\inst{\ref{inst12}}, 
G. Yepes\inst{\ref{inst44}}, 
S. Zarattini\inst{\ref{inst1},\ref{inst46}}
}
\authorrunning{The CHEX-MATE Collaboration}

\institute{AIM, CEA, CNRS, Université Paris-Saclay, Université Paris Diderot, Sorbonne Paris Cité, F-91191 Gif-sur-Yvette, France \email{marn.sett@gmail.com}
\label{inst1} \\
\and
INAF, Osservatorio di Astrofisica e Scienza dello Spazio, via Pietro Gobetti 93/3, 40129 Bologna, Italy \email{marn.sett@gmail.com}
\label{inst2} \\
\and
INFN, Sezione di Bologna, viale Berti Pichat 6/2, I-40127 Bologna, Italy
\label{inst3} \\
\and
INAF, IASF-Milano, via A. Corti 12, I-20133 Milano, Italy
\label{inst4} \\
\and
Department of Astronomy, University of Geneva, ch. d’\'Ecogia 16, CH-1290 Versoix Switzerland
\label{inst5} \\
\and
CNRS and Sorbonne Universit\'e, UMR 7095, Institut d’Astrophysique de Paris, 98 bis Boulevard Arago, 75014 Paris, France
\label{inst6} \\
\and
Jodrell Bank Centre for Astrophysics, Department of Physics and Astronomy, School of Natural Sciences, The University of Manchester, Manchester M13 9PL, UK
\label{inst7} \\
\and
Center for Astrophysics $|$ Harvard $\&$ Smithsonian, 60 Garden Street, Cambridge, MA 02138, USA
\label{inst8} \\
\and
HH Wills Physics Laboratory, University of Bristol, Tyndall Ave, Bristol, BS8 1TL, UK
\label{inst9} \\
\and
IRAP, Universit\'e de Toulouse, CNRS, CNES, UPS, 9 av du colonel Roche, BP44346, 31028 Toulouse cedex 4, France
\label{inst10} \\
\and
Dipartimento di Fisica e Astronomia, Universit\'{a} di Bologna, Via Gobetti 92/3, 40121, Bologna, Italy
\label{inst11} \\
\and
INAF, Istituto di Radio Astronomia, Via Gobetti 101, 40129 Bologna, Italy
\label{inst12} \\
\and
Hamburger Sternwarte, Gojenbergsweg 112, 21029 Hamburg, Germany
\label{inst13} \\
\and
Universit\`a degli studi di Roma “Tor Vergata”, via della ricerca scientifica, 1, 00133 Roma, Italy
\label{inst14} \\
\and
Universit\'e C\^{o}te d’Azur, Observatoire de la C\^{o}te d’Azur, CNRS, Laboratoire Lagrange, Nice, France
\label{inst15} \\
\and
INAF – Osservatorio Astronomico di Trieste, via Tiepolo 11, I-34131 Trieste, Italy
\label{inst16} \\
\and 
IFPU – Institute for Fundamental Physics of the Universe, via Beirut 2, 34151, Trieste, Italy
\label{inst17} \\
\and
California Institute of Technology, Pasadena, CA 91125, USA
\label{inst18} \\
\and
Department of Physics and Kavli Institute for Astrophysics and Space Research, Massachusetts Institute of Technology, Cambridge, MA 02139, USA
\label{inst19} \\
\and
Instituto de Astrofísica de Canarias (IAC), C/ Vía Láctea s/n, E-38205, La Laguna, Tenerife, Spain
\label{inst20} \\
\and 
Universidad de La Laguna, Departamento de Astrofísica, C/ Astrofísico Francisco Sánchez s/n, E-38206, La Laguna, Tenerife, Spain
\label{inst21} \\
\and
Dipartimento di Fisica, Sezione di Astronomia, Università di Trieste, Via Tiepolo 11, I-34143 Trieste, Italy
\label{inst22} \\
\and
INFN – Sezione di Trieste, I-34100 Trieste, Italy
\label{inst23} \\
\and
LUTH, UMR 8102 CNRS, Observatoire de Paris, PSL Research University, Universit\'e of Paris, 5 place Jules Janssen, 92195 Meudon, France
\label{inst24} \\
\and
Sorbonne Universit\'e, CNRS, UMR 7095, Institut d’Astrophysique de Paris, 98 bis bd Arago, 75014 Paris, France
\label{inst25} \\
\and
INAF,  Osservatorio Astronomico di Brera, via E. Bianchi 46, 23807 Merate, Italy
\label{inst26} \\
\and
Dipartimento di Fisica, Sapienza Universit\'a di Roma, Piazzale Aldo Moro, 5-00185 Roma, Italy
\label{inst27} \\
\and
University Observatory Munich, Scheinerstrasse 1, 81679 Munich, Germany
\label{inst28} \\
\and 
Max-Planck-Institut fuer Astrophysik, Karl-Schwarzschild-Str. 1, D-85748 Garching, Germany
\label{inst29} \\
\and
Physics and Astronomy Department, Michigan State University, East Lansing, MI, 48824, USA
\label{inst30} \\
\and
Department of Astrophysical Sciences, Princeton University, 4 Ivy Lane, Princeton, NJ 08544-1001, USA
\label{inst31} \\
\and
Centre for Extragalactic Astronomy, Durham University, South Road, Durham DH1 3LE, UK
\label{inst32} \\
\and 
Institute for Computational Cosmology, Durham University, South Road, Durham DH1 3LE, UK
\label{inst33} \\
\and 
Astrophysics Research Centre, University of KwaZulu-Natal, Westville Campus, Durban 4041, South Africa
\label{inst34} \\
\and 
School of Mathematics, Statistics \& Computer Science, University of KwaZulu-Natal, Westville Campus, Durban 4041, South Africa
\label{inst35} \\
\and
Curtin Institute for Computation, GPO Box U1987, Perth, WA 6845, Australia
\label{inst36} \\
\and
International Centre for Radio Astronomy Research (ICRAR), Curtin University, Bentley, WA 6102, Australia
\label{inst37} \\
\and
Univ. Grenoble Alpes, CNRS, LPSC-IN2P3, 53, avenue des Martyrs, 38000 Grenoble, France
\label{inst38} \\
\and
IRFU, CEA, Universit{\'e} Paris-Saclay, F-91191 Gif-sur-Yvette, France
\label{inst39} \\
\and
Physics Program, Graduate School of Advanced Science and Engineering, Hiroshima University, 1-3-1 Kagamiyama, Higashi-Hiroshima, Hiroshima 739-8526, Japan
\label{inst40} \\
\and
INAF, Osservatorio Astrofisico di Arcetri, Largo E. Fermi 5, I-50125, Firenze, Italy
\label{inst41} \\
\and
Institute of Astronomy and Astrophysics, Academia Sinica, No. 1, Section 4, Roosevelt Road, Taipei 10617
\label{inst42} \\
\and
European Southern Observatory, Karl-Schwarzschild-Str. 2, 85748, Garching, Germany
\label{inst43} \\
\and
Departamento de Física Teórica and CIAFF,  Facultad de Ciencias, Modulo 8, Universidad Autónoma de Madrid, 28049, Madrid, Spain
\label{inst44} \\
\and 
Instituto de Astronom\'{i}a y Ciencias Planetarias de Atacama (INCT), Universidad de Atacama, Copayapu 485, Copiap\'{o}, Chile
\label{inst45} \\
\and 
Dipartimento di Fisica e Astronomia “G. Galilei”, Universit\'{a}  di Padova, vicolo dell’Osservatorio 3, I-35122 Padova, Italy \label{inst46} \\
\and
INAF – Osservatorio Astronomico di Padova, vicolo dell’Osservatorio 5, I-35122 Padova, Italy  \label{inst47} \\
\and
Observatorio Astron\'omico Nacional (OAN-IGN), C/ Alfonso XII 3, E-28014, Madrid, Spain \label{inst48} \\
}

%%%%%%%%%%%%%%%%%%

\abstract
{The Cluster HEritage project with XMM-\textit{Newton} - Mass Assembly and Thermodynamics at the Endpoint of structure formation (CHEX-MATE) is a three-mega-second Multi-Year Heritage Programme to obtain X-ray observations of a minimally-biased, signal-to-noise-limited sample of 118 galaxy clusters detected by \planck\ through the Sunyaev-Zeldovich effect. The programme, described in detail in this paper, aims to study the ultimate products of structure formation in time and mass. It is composed of a census of the most recent objects to have formed (Tier-1: $0.05 < z < 0.2$; $2\times10^{14}$ M$_{\odot} < \Mv < 9 \times 10^{14}$ M$_{\odot}$), together with a sample of the highest mass objects in the Universe (Tier-2: $z < 0.6$; $\Mv > 7.25 \times 10^{14}$ M$_{\odot}$). The programme will yield an accurate vision of the statistical properties of the underlying population, measure how the gas properties are shaped by collapse into the dark matter halo, uncover the provenance of non-gravitational heating, and resolve the major uncertainties in mass determination that limit the use of clusters for cosmological parameter estimation. We will acquire X-ray exposures of uniform depth, designed to obtain individual mass measurements accurate to $15-20\%$ under the hydrostatic assumption. We present the project motivations, describe the programme definition, and detail the ongoing multi-wavelength observational (lensing, SZ, radio) and theoretical effort that is being deployed in support of the project. }


\maketitle 

%%%%%%%%%%%%%%%%%%

\section{Introduction}

Clusters of galaxies provide valuable information on cosmology, from the physics driving galaxy and structure formation, to the nature of dark matter and dark energy \citep[see e.g.][]{all11,kb12}.
They are the nodes of the cosmic web, constantly growing through accretion of matter along filaments and via occasional mergers, and their matter
content reflects that  of the Universe ($\sim85\%$ dark matter, $\sim12 \%$ X-ray emitting gas and $\sim3\%$ galaxies). Clusters are therefore excellent laboratories for probing the physics of the gravitational collapse of dark matter and baryons, 
and for studying the non-gravitational physics that affects their baryonic component. As cluster growth and evolution depend on the underlying cosmology 
(through initial conditions, cosmic expansion rate, and dark matter properties), their number density as a function of mass and redshift, their spatial distribution, and their internal structure, are powerful cosmological probes.

Historically, optical and X--ray surveys have been the primary source of cluster catalogues.
However, they can also be detected and studied via the Sunyaev-Zel'dovich effect \citep[SZE; ][]{sz72,bir99,car02,mro19}, the spectral distortion of the cosmic microwave background (CMB) generated through inverse Compton scattering of CMB photons by the hot electrons in the intra-cluster medium (ICM). 
The SZE brightness is independent of the distance to the object, and the total signal, $\YSZ$, is proportional to the thermal energy content of the ICM and is expected to be tightly correlated to the total mass \citep{das04,mot05}. SZE surveys such as those from the Atacama Cosmology Telescope \citep[ACT;][]{mar11,has13,hil18}, the South Pole Telescope  \citep[SPT;][]{ble15,ble20} and \planck\ \citep{esz,psz1,psz2} have provided cluster samples up to high $z$. These are thought to be as near as possible to being mass-selected, and as such are minimally-biased. 
The advent of these SZE-selected cluster catalogues, combined with new and archival X--ray information, has been transformational.

Indeed, X-ray follow-up of these new objects has raised new questions. The discovery that X-ray-selected and SZ-selected samples do not appear to have the same distribution of dynamical states \citep[e.g.][]{planck11_9,ros16,and17,lov17} has prompted examination of the relationship between the baryon signatures and the underlying cluster population. The possible tension between the value of the normalised matter density parameter, $\sigma_8$, obtained from cluster number counts, and that based on CMB measurements, has stimulated work on the absolute mass scale of clusters \citep[see][for a review]{pra19}. With the advent of very high redshift ($z > 1$) detections through the SZE,  the issue of how to build a fully consistent picture of population evolution has also come to the fore.

In its 2017 Announcement of Opportunity, ESA offered the possibility to propose Multi-Year Heritage (MYH) programmes using the X-ray satellite \xmm\ for the first time. This prompted a large response from the community, and the resulting oversubscription factor for the MYH proposal class was around ten. Two programmes were awarded  MYH status, one of which was led by our collaboration  (PIs: M. Arnaud, CEA Saclay; S. Ettori, INAF OAS Bologna). The project, now titled  Cluster HEritage project with XMM-\textit{Newton} - Mass Assembly and Thermodynamics at the Endpoint of structure formation (CHEX-MATE)\footnote{\url{xmm-heritage.oas.inaf.it}}, aims to obtain complete and homogeneous X-ray exposures of 118 \planck\ SZE-selected galaxy clusters. 

The sample comprises a minimally biased census both of the population of clusters at the most recent time ($0.05<z<0.2$), and of the most massive objects to have formed at $z<0.6$. It was designed to answer the following questions: 
\begin{itemize}

\item What is the absolute cluster mass scale? What is the imprint of the formation process on the equilibrium state of clusters, and how does this impact our ability to weigh them through their baryon signature? 

\item What are the statistical properties of the cluster population? How does cluster detectability depend on baryon physics?

\item Can we accurately measure how the properties of the cluster population change over time? What are the ultimate products of structure formation?

\end{itemize}

The X-ray observations started in Summer 2018 and would continue for three years. All the data are made public as soon as they are acquired. \planck\ SZE data are available for the full data set. Access to independent weak lensing (WL) mass estimates for a sizeable fraction of the sample was ensured through an  object selection strategy that optimised coverage with existing high-quality optical multi-band imaging data from the Canada-France-Hawaii Telescope (CFHT), Subaru, and the {\it Hubble} Space Telescope (HST). Coverage was also optimised with the ongoing Canada-France Imaging Survey (CFIS) on the CFHT, and with the {\it Euclid} survey footprints. 

This paper presents a detailed overview of the project, and acts as a reference for the collaboration and for the wider community. It is organised as follows. 
In Sect.~\ref{sec:xhcp}, we present the motivating questions, scientific goals, and legacy value of such a project. In Sect.~\ref{sec:strat}, we discuss the sample definition and observation strategy. Section~\ref{sec:support} describes the supporting data essential to the project goals, and we present our summary and conclusions in Sect.~\ref{sec:conc}. Throughout the paper we assume a flat $\Lambda$CDM cosmology with $\Omega_m=0.3$, $\Omega_\Lambda=0.7$ and $H_{0}=70$ km s$^{-1}$ Mpc$^{-1}$. The variables $M_{\Delta}$ and $R_{\Delta}$ are the total mass and radius corresponding to a total density contrast $\Delta \, \rho_{\rm c}(z)$, where $\rho_{\rm c}(z)$ is the critical density of the Universe at the cluster redshift; thus, for example, $M_{500} = (4\pi/3)\,500\,\rho_{\rm c}(z)\,R_{500}^3$.  The quantity $Y_{\rm X}$ is the product of $\Mgv$, the gas mass within $\Rv$, and $\TX$, the spectroscopic temperature measured in the $[0.15$--$0.75]~\Rv$ aperture. The SZ flux is characterised by $Y_{\Delta}$, where $Y_{\Delta}\,D_{\rm A}^2$ is the spherically integrated Compton parameter within $R_{\Delta}$, and $D_{\rm A}$ is the angular-diameter distance to the cluster.

%%%%%%%%%%%%%%%%%%

\section{CHEX-MATE description}
\label{sec:xhcp}

%%-----------
% z-M figure
\begin{figure*}
  \begin{minipage}[t]{0.65\textwidth}
  \vspace{0pt}
    \includegraphics[width=\textwidth]{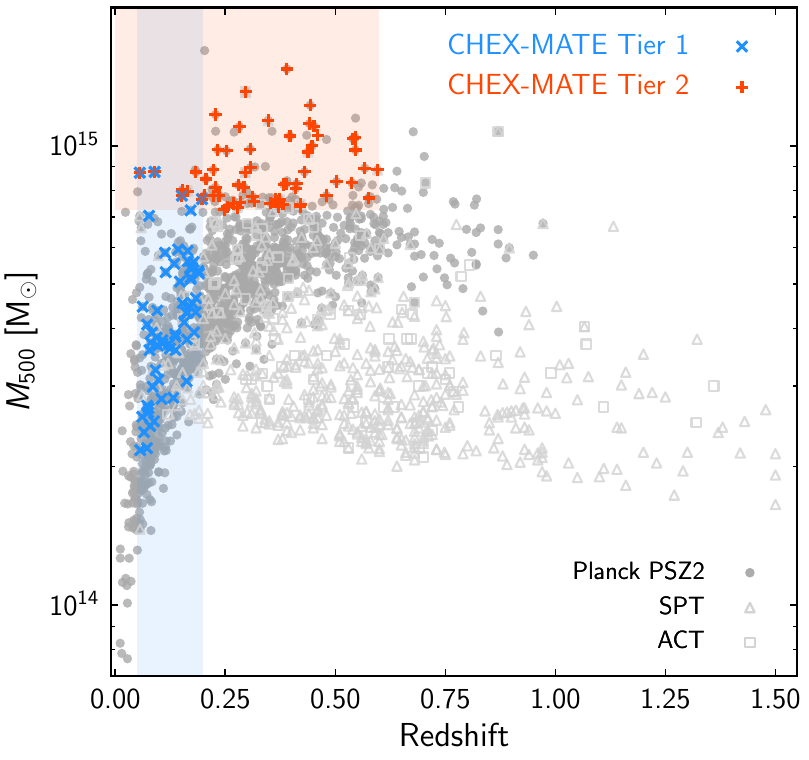}
  \end{minipage}\hfill
  \begin{minipage}[t]{0.3\textwidth}
  \vspace{0pt}
    \caption{
\footnotesize Distribution in the $M_{\rm 500}-z$ plane of confirmed clusters from major SZE surveys,  available at the time of proposal submission (October 2017). Filled circles: \planck\  clusters  \citep{esz,psz1,psz2}; triangles: SPT \citep{ble15}; squares: ACT \citep{has13}. Masses for \planck\ clusters are derived iteratively from the $\YSZ$--$\Mv$ relation calibrated using masses from \xmm; these were not corrected for any HE bias (see text for details). The figure includes both  masses published in the \planck\ catalogue, and new masses computed  using new redshift information. The shaded boxes indicate the Tier-1 and Tier-2 redshift ranges in blue and orange, respectively. The sample is drawn from the \planck\ PSZ2 sample, selecting clusters detected at high signal-to-noise ratio (S/N$>6.5$) with the MMF3 algorithm, and in the cleanest part of the sky. We also excluded clusters in the sky region with poor \xmm\ visibility. Additional redshift, sky area, mass criteria,  are applied to define the Tier-1 ($0.05<z<0.2;\ {\rm Dec} >0$)  and Tier-2  ($z<0.6$, $M_{500}>7.25 \times 10^{14}\msol$) samples. 
Remaining clusters in the shaded part of the $M_{\rm 500}-z$ plane are at lower S/N, or lie outside the sky regions under consideration. A full description of the sample strategy is given in Sec.~\ref{sec:sample} and is further illustrated in Appendix~\ref{sec:appx}.
} \label{fig:zM}
  \end{minipage}
\end{figure*}
%-----------

%%%%%%%%%%%%%%%%%%

\subsection{Motivating questions}

Inspired by the new results obtained from the objects found in SZE-selected cluster surveys, and from their subsequent multi-wavelength follow-up, the project is built around a series of questions. 

\subsubsection{What is the absolute cluster mass scale?}

Theory predicts the number of clusters as a function of their redshift and mass. Surveys detect clusters through their observable baryon signature such as their X-ray or SZE signal, or the optical richness. To obtain cosmological constraints from the cluster population, this signal must then be linked to the underlying mass; in other words, one must know the relation between the observable and the mass, and the scatter about this relation. One must also understand the probability that a cluster of a given mass is detected with a given value of the survey observable; the resulting selection function is a key element in the cosmological analysis of the cluster population. 

In the first \planck\ SZE cluster cosmology analysis, the SZE-mass scaling relation was derived from X‐ray observations and numerical simulations. They combined the $M_{500}$–-$Y_{X}$ relation obtained from a sample of relaxed clusters with masses derived from the hydrostatic equilibrium (HE) equation \citet{arn10}, and the $Y_{\rm X}$–-$Y_{\rm SZ}$ relation calibrated on a subset of clusters from the cosmology sample \citep[][Appendix A]{pla13szcosmo}. They introduced a mass bias parameter, $b$, to account for differences between the X-ray mass estimates and the true cluster halo mass: $M_\Delta = (1-b)\, M_{\Delta, {\rm true}}$. The factor $b$ encompasses all unknowns with regard to the relationship between the X-ray mass and the true mass, such as can arise from observational effects such as instrumental calibration, or from cluster physics such as departure from HE or temperature structure in the ICM.

The main result from the \planck\ SZE cluster count  analysis was that, with a fiducial $(1-b) = 0.8$, derived from numerical simulations, the $\sigma_8$ and $\Omega_{\rm m}$  values obtained from SZE cluster abundances were inconsistent at the $\sim 2\sigma$ level  with the values derived from the \planck\ CMB cosmology \citep{planck16_sz, planck16_cosmo}. For the 2015 analysis, a value of $(1-b) = 0.58 \pm 0.04$ would be needed  reconcile cluster counts and CMB measurements, implying a much larger HE bias than expected from numerical simulations. The value needed to reconcile cluster counts and CMB reduces to $(1-b)=0.62\pm0.03$ in the 2018 \planck\ CMB analysis. This is still considerably larger than expectations. Inclusion of additional constraints from the thermal SZ power spectrum similarly implies $(1-b) \lesssim 0.67$ \citep{sal18}. 

Prompted by these results, the cluster mass determination, and its relation to 
the observable, have become issues of great debate in the community \citep[see e.g. the review of][]{pra19}. Important new constraints on the value of $(1-b)$ have come from WL mass measurements of sizeable samples with good control of systematic effects  (e.g. the Cluster Lensing and Supernova Survey with Hubble - CLASH, \citealt{pos12}; the Canadian Cluster Cosmology Project - CCCP, \citealt{hoe15,her+al19}; Weighing the Giants - WtG, \citealt{vdl14}; the Local Cluster Substructure Survey -  LoCuSS, \citealt{smi16};  PSZ2LenS, \citealt{ser+al17_psz2lens}). However, a consensus has not been reached, with, for example, WtG finding $(1-b) = 0.69 \pm 0.07$, 
marginally reconciling CMB and cluster constraints \citep{planck16_cosmo} and implying a large HE bias, but LoCuSS measuring $(1-b) = 0.95\pm0.04$, indicating a low HE bias. An alternative mass measurement from lensing of the CMB itself by clusters initially suggested no significant bias  \citep[e.g.][]{mel15}; however, recent re-analysis by \citet{zub19}, including the mass bias factor directly in the cosmological analysis, finds $(1-b) = 0.71 \pm 0.10$.

The theoretical picture is also uncertain. A significant upward revision of the total mass would imply that cluster baryon fractions were  
significantly lower than the universal value, at odds with expectations from numerical simulations \citep[e.g.][]{pla17,ans20}.  Similarly, while  simulations predict some turbulence and non-thermal pressure support from gas motions generated by the hierarchical assembly process, they  do not indicate that clusters are strongly out of equilibrium on average \citep[e.g.][]{bif16,ans20,ang20}. 
Recent observational constraints also suggest that this is not the case, at least in relaxed nearby massive systems \citep{eck19}.

Larger samples of high-quality data are needed to reduce the statistical uncertainties in the absolute mass calibration, and to fully characterise any residual intrinsic scatter. This can best be achieved through a sample selection strategy that reflects as closely as possible the underlying population.

%%%%%%%%%%%%%%%%%%

\subsubsection{What is the `true' underlying cluster population?}

Current surveys detect clusters through their baryon signature. The SZE signal, proportional to the integral of the gas pressure along the line of sight,
has been shown to behave well,  with a weak dependence on dynamical state and on poorly understood non-gravitational  physics \citep{das04,pla17}. 
A comparison of \planck\ SZE selected clusters with X-ray selected clusters indicated that the former are on average less relaxed 
(using gas morphological indicators or BCG-centre offset), and contain a lower fraction of over-dense, cool core systems \citep[][see also \citealt{zen20} for a different view]{planck11_9,ros16,ros17,and17,lov17}.  

This may reflect the tendency of X-ray surveys to preferentially detect clusters with a centrally-peaked  morphology, which are  more luminous at a given mass, 
and on average more relaxed \citep[e.g.][]{pes90,pac07,eck11}. 
However,  it is currently unclear if this selection effect is sufficient to explain the difference \citep[e.g.][]{ros17}.
This also raises concerns about how
representative the X-ray selected samples,  used to define our current understanding of cluster physics and to calibrate numerical simulations, have been. Examples, frequently used in the literature,  include  the \rexcess\ sample of 33 clusters with deep \xmm\ data \citep{boe07,pra09,pra10,arn10}, or the sample of relaxed clusters with deep \cxo\ observations  studied by \citet{vik06}.

We expect a sample selected through its SZE signal to be more representative of the underlying population, and as such the least biased that it is currently possible to obtain.
The ensemble properties of such a sample will yield critical insights into the gas thermodynamic properties and their relation to the cluster mass, and into how variations in gas properties feed into the survey selection function.

%%%%%%%%%%%%%%%%%%

\subsubsection{Can we measure how the properties of the cluster population change over time?}

\cxo\ follow-up of clusters detected by the SPT between redshift $0.3$ and $1.9$ has indicated 
that the average ICM properties outside the core are remarkably self-similar,  
with no measurable evolution of morphological dynamical indicators  \citep{mcd14,mcd17,nur17}. These observations also suggested that cool cores are formed early and are very stable to further dynamical evolution. However, as the SPT survey is highly incomplete below $z=0.3$, this study relies on an X-ray-selected sample to provide the low-$z$ anchor. Due to the selection effects outlined above, we do not yet have a fully consistent picture of population evolution. 

The redshift independence of the SZE has led to the discovery of many hundreds of high-redshift systems, with which studies of how the properties of the  cluster population change with time can be undertaken. However, such studies need a well-characterised low-redshift anchor obtained with the same selection method.

%%%%%%%%%%%%%%%%%%

\subsection{Immediate scientific goals}

The questions discussed above led to the definition of CHEX-MATE, a sample of 118 clusters detected by \planck\ at high signal-to-noise (S/N$>6.5$) through their SZE signal. Figure~\ref{fig:zM} shows the sample in the $z-M$ plane. It is composed of:

\begin{itemize}
    \item Tier-1: a census of the population of clusters at the most recent time ($0.05 < z < 0.2$, with $2 \times 10^{14}$ M$_{\odot} < \Mv < 9 \times 10^{14}$ M$_{\odot}$);
    \item Tier-2: the most massive systems to have formed thus far in the history of the Universe ($z < 0.6$,  with $\Mv > 7.25 \times 10^{14}$ M$_{\odot}$).
\end{itemize}

\noindent The 61 clusters in Tier-1 provide an unbiased view of the population at the present time, and serve as the fundamental anchor of any study that seeks to assess how the population changes over cosmic time.  
        The 61 objects in Tier-2 comprise the most massive clusters, the ultimate manifestation of hierarchical structure formation, which the local volume is too limited to contain. Four systems are common to both Tiers. In the following, we describe the detailed scientific goals of the project.

%%%%%%%%%%%%%%%%%%

\subsubsection{The dynamical collapse of the ICM}

The extent to which the gas is in equilibrium in the dark matter potential, as a function of mass and radius, is a key issue for the understanding of the mass scale.  
This is linked to the presence of turbulence in the ICM, non-thermal electrons (detectable in radio emission), shocks, bulk motion, and sub-clustering at all scales. 
Objective morphological indicators (e.g. centroid shifts, power ratios etc) will be provided by the X--ray imaging \citep{lov17}. 
An exciting new development is the use of surface brightness fluctuations to constrain the turbulence spectrum \citep{gas13,zhur14,hof15,eck17}. 
Combining SZE and X-ray imagery will allow us to constrain gas clumpiness and the thermodynamical properties in the outskirts, as addressed in the X-COP project (see e.g. \citealt{eck17_xcop,ghi19_univ,eck19,ett19}). 
We will measure various key ICM parameters, their dependence on mass, and study outliers in detail.  
These results will provide key information for our investigation of mass biases, as discussed below. 
We will correlate with radio surveys to link the dynamical indicators to the presence and extent  of  non-thermal energy contained in radio halos and relics.

Furthermore, simulations show that the most massive clusters always form at the crossroads of the hottest filaments. Objects with $M \sim 10^{15}~ \rm M_{\odot}$ have an   $\geq 80\%$ probability of being connected  by a filament of dark and luminous matter to a neighbouring cluster at a distance of $<15\ {\rm Mpc}/h$ \citep{col05}. The field of view (FoV) of \xmm\ allows the study of the large-scale environment of massive clusters, since a single pointing is sufficient to map the entire azimuth above $R_{200}$ in most of the massive (Tier-2) objects. In particular, in more than 60\% of the Tier-2 objects, the \xmm\ FoV subtends a region up to 2$R_{200}$.
These systems are the ideal targets for a robust detection of the large-scale cosmic web \citep[e.g.][]{hai18}. The possibility of studying gas compression and dynamical activity between clusters in an early merger stage has recently been raised by several radio observations \citep[e.g.][]{aka17,gov19,bot20} and in numerical simulations \citep[e.g.][]{vaz19}. 
Detecting and studying the rare merger configurations that may lead to the formation of cluster-cluster bridges will be an additional challenge for CHEX-MATE.

%%%%%%%%%%%%%%%%%%

\subsubsection{The cluster mass scale}

\begin{figure*}[t]
\centering
\includegraphics[width=0.85\textwidth]{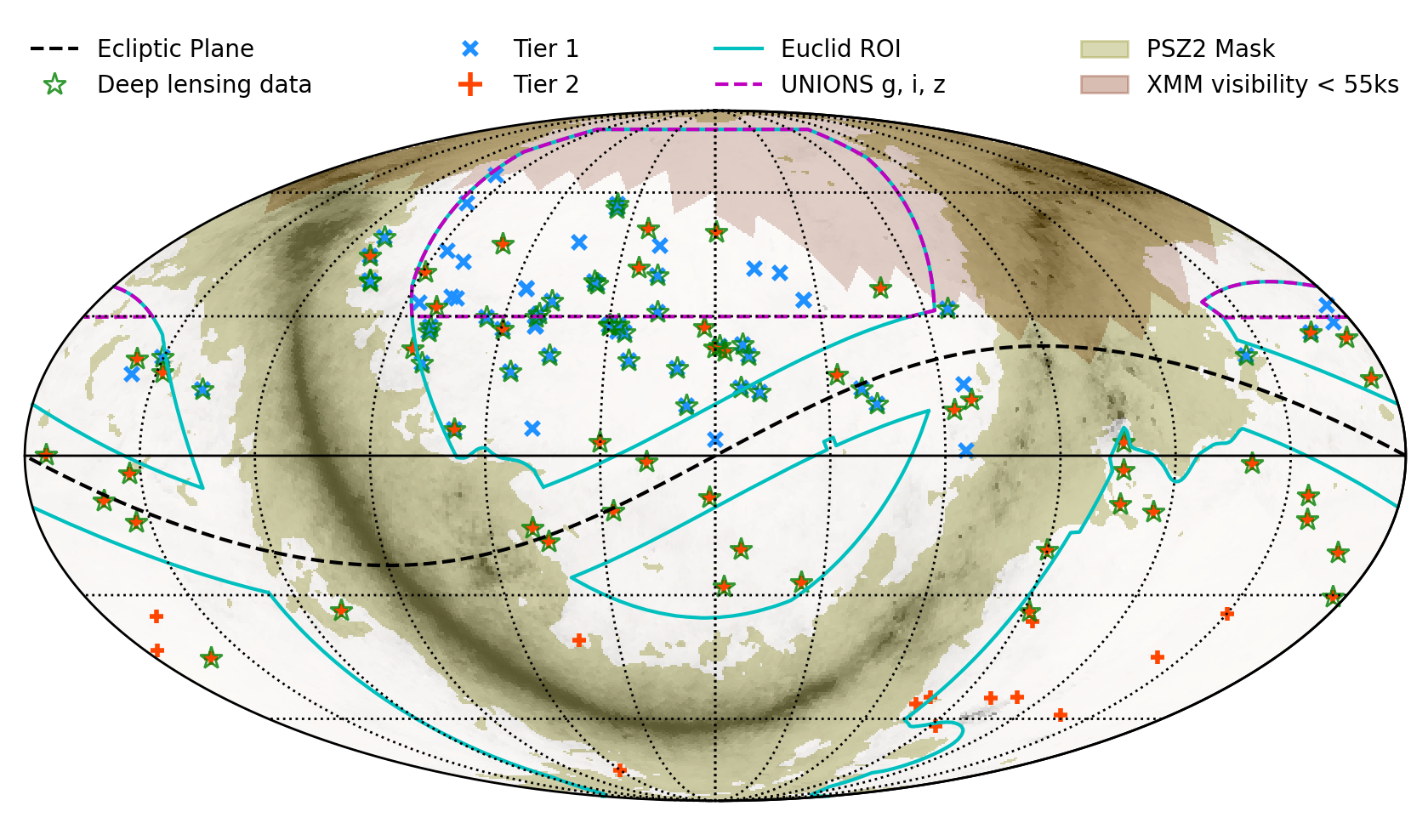}
\caption{\footnotesize Sky distribution of Tier-1 (green) and Tier-2 (red) clusters detected  in the \planck\ `cosmological' mask, outside the dusty galactic plane (green shaded area),  and outside the region of low \xmm\ visibility (red shaded area). A further cut in declination is applied for Tier-1. The synergy in sky coverage between our SZE selected sample and the UNIONS/CFIS/{\it Euclid} lensing surveys is clear. In total, 107 targets fall in the footprint of the {\it Euclid} Region of Interest survey (cyan) and  33 in the UNIONS/CFIS survey footprint (pink).
Clusters with existing or ongoing deep WL mass measurements are marked by stars (see Sect.~\ref{sec:lens} for details).
}
\label{fig:lens}
\end{figure*}

We will measure total integrated mass profiles (out to at least $\Rv$) for all objects using the equations derived from the HE assumption \citep[e.g.][]{pra02,ett13}. 
The total HE mass will be compared to mass proxies such as the SZE signal $\YSZ$, the X-ray luminosity $\LX$ or  $\YX$ (the product of ICM mass and temperature). Most importantly,  WL data are already available for a significant fraction of the  sample, especially at high mass (see Fig.~\ref{fig:lens}). Section~\ref{sec:lens} details the currently available lensing data and details the strategy we have deployed to obtain complete WL follow-up. Ultimately, follow-up will also be available with {\it Euclid}\footnote{{\it Euclid}: \url{sci.esa.int/web/euclid}}.

Comparison of these  mass estimates (weak lensing mass M$_{\rm WL}$, hydrostatic mass M$_{\rm HE}$) and various mass proxies can be undertaken, measuring the best fitting scaling laws and scatter, and the covariance between quantities. Correlation with dynamical indicators and  investigation of trends with mass can also be performed.  
This will be the first time that such an investigation of cluster masses will be performed systematically and self-consistently on a well defined and minimally-biased sample, covering the full mass range.  Many comparisons based on reference samples (e.g. the \planck\ calibration samples, LoCuSS, CCCP, WtG) yield only a partial overview of the inter-dependence of the parameters (e.g. $M_{\rm WL}$--$M_{\rm HE}$ or $M_{\rm WL}$--$\YSZ$),  as they are statistically incomplete due to limited coverage, or were compiled based on 
criteria such as archival availability.

 All mass estimates are subject to inherent bias (see e.g. the review by \citealt{pra19} and references therein). The HE bias is well known to affect X-ray observations, but lensing is also subject to biases due to line-of-sight effects. While the lensing mass is expected to be the least biased {\it on average}, it is of lower statistical quality on an individual cluster basis \citep[e.g.~][]{Men++10,Ham++12}.
 Our goal  is to  build a consistent  understanding  of  the various biases and to define  the best strategy to obtain the most accurate  mass estimate  in various surveys. 

%%%%%%%%%%%%%%%%%%

\subsubsection{The interplay between gravitational and non-gravitational processes}

The  densest core regions, where the interplay between cooling and central AGN feedback is strongest, provide key diagnostics on the impact of non-gravitational processes on the ICM \citep[e.g.][]{cav09,pra10}.  
If cool cores are less prominent than previously thought from X-ray selected samples, we may have to fundamentally revise our vision of cooling and galaxy feedback at cluster scales. 

With this sample, the true distribution of cool core strength \citep[see e.g.][]{hud10} can be reassessed, as can the impact of feedback on the thermodynamical properties of the ICM as a function of radius, mass, and, at the high mass end, redshift. 
We can definitively establish the relation between core properties and the bulk, including  dynamical state (e.g. Are cool cores essentially found in relaxed systems? To what extent are they destroyed by mergers?), thereby providing a testbed for predictions from numerical simulations \citep[see e.g.][]{bar18a}.

As shown by a diverse range of studies, linking AGN feeding and feedback processes over nine orders of magnitude is vital to advancing our understanding of clusters and diffuse hot halos (see  e.g. \citealt{mcd18}, \citealt{gas20} for reviews). We can establish the new population-level baseline to understand the interplay between gravitational heating, cooling and AGN feedback. Covering the full range of masses probed by \planck, the sample includes both the highest mass systems dominated  by gravitational heating, and lower mass systems that are  progressively more affected by non-gravitational input.
The radial coverage, from the core to at least $\Rv$,  is equally important for  sampling the relative impact of the different energetic processes, and for obtaining the widest possible view of the gas morphology.

The measurement of metal abundances in the ICM is a powerful probe of the nature of galaxy feedback processes (see \citealt{mern18}, \citealt{bif18}, and references therein). The abundances yield information both on the various types of supernovae (core-collapse and SNIa) producing the metals throughout the cluster lifetime (reaching back to the proto-cluster phase), and on the AGN feedback mechanisms that spread the metals throughout the ICM. 
Although not tailored to the measurement of metal abundances out to $\Rv$, our observations will enable measurement of the total amount of iron out to a significant fraction of $\Rv$. We can test the uniformity of the metal enrichment in massive clusters as a function of redshift with Tier-2, and as a function of mass with Tier-1. By comparing with stellar masses, we will address the long-standing issue of whether the amount of iron in the ICM is in excess of what can be produced in the stars \citep[e.g.][]{arn92,ghi20}; and in particular with Tier-1, address the relation of the iron mass, ICM mass and stellar mass, to the total mass \citep[e.g.][]{breg10,renz14}.

%%%%%%%%%%%%%%%%%%

%------------
% Sky location figure
%------------

%%%%%%%%%%%%%%%%%%

\subsubsection{A local anchor for tracking population changes}

Our project will yield the ultimate baseline for  the statistical properties of nearby clusters and of the most massive clusters to have formed in $5.8$\,Gyr look back time. It is based on a sample defined to be as unbiased as possible for detection based on baryon observables. We emphasise that the X-ray and lensing properties that we intend to measure will be independent of the detection signal, 
minimising the need for Eddington bias correction (although covariances between quantities will need to be taken into account).
The major outputs of our project will include scaling laws, structural properties, and quantitative dynamical indicators, including dispersion and covariance between parameters.
Tier-1 has three times more  clusters than \rexcess,  permitting  a major step forward on the precision not only of the main trends, but also of the dispersion around them. The full sample size and mass coverage will allow the dispersion to be explored as a function of mass, and, at high mass, also as a function of redshift. 
Crucially, this work will be underpinned by the  best possible control of systematics on cluster masses due to our self-consistent study of the mass scale and related biases. Our work will provide a state-of-the-art reference with which to anchor our view of how the population changes with time from ongoing  \cxo\ and \xmm\ follow-up of high-$z$ SZE clusters, and with which to calibrate the baryon physics in numerical simulations that are used to interpret surveys (e.g. as undertaken in the BAHAMAS project by \citealt{mcc17}; see also \citealt{ras15} and the discussion in Sect.~\ref{sec:sims}).

The project is of substantial value for next-generation X-ray and SZE surveys. Our sample corresponds to the descendants of the high--$z$ objects that will be detected by upcoming SZE surveys such as SPT-3G, which will probe lower masses than currently possible, and as such represents the culmination of the cluster  evolutionary track. The project will also provide key input  for the interpretation of eROSITA\footnote{eROSITA: extended Röntgen Survey with an Imaging Telescope Array: \url{www.mpe.mpg.de/eROSITA}}, the ongoing All-sky X-ray survey. The X-ray luminosity depends on the square of the gas density and  is dominated by the core properties, which presents a large scatter and a strong dependence on thermodynamical state and the effect of non-gravitational processes. 
X--ray cluster detectability further depends on morphology, which is closely linked to the dynamical state \citep[see Fig. 2 in][]{arn17}. We can investigate the X-ray luminosity--mass relation and its scatter, together with its relation to the  distribution of morphologies in the population, enabling us to understand these selection effects. Combined with improved measurements of cluster evolution, our work will provide the basis for robust modelling of the selection for any X-ray survey. 

% --------------
\begin{figure*}[t]
\centering
\includegraphics[bb=60 340 550 785 ,clip,scale=1.,angle=0,keepaspectratio,width=0.90\textwidth]{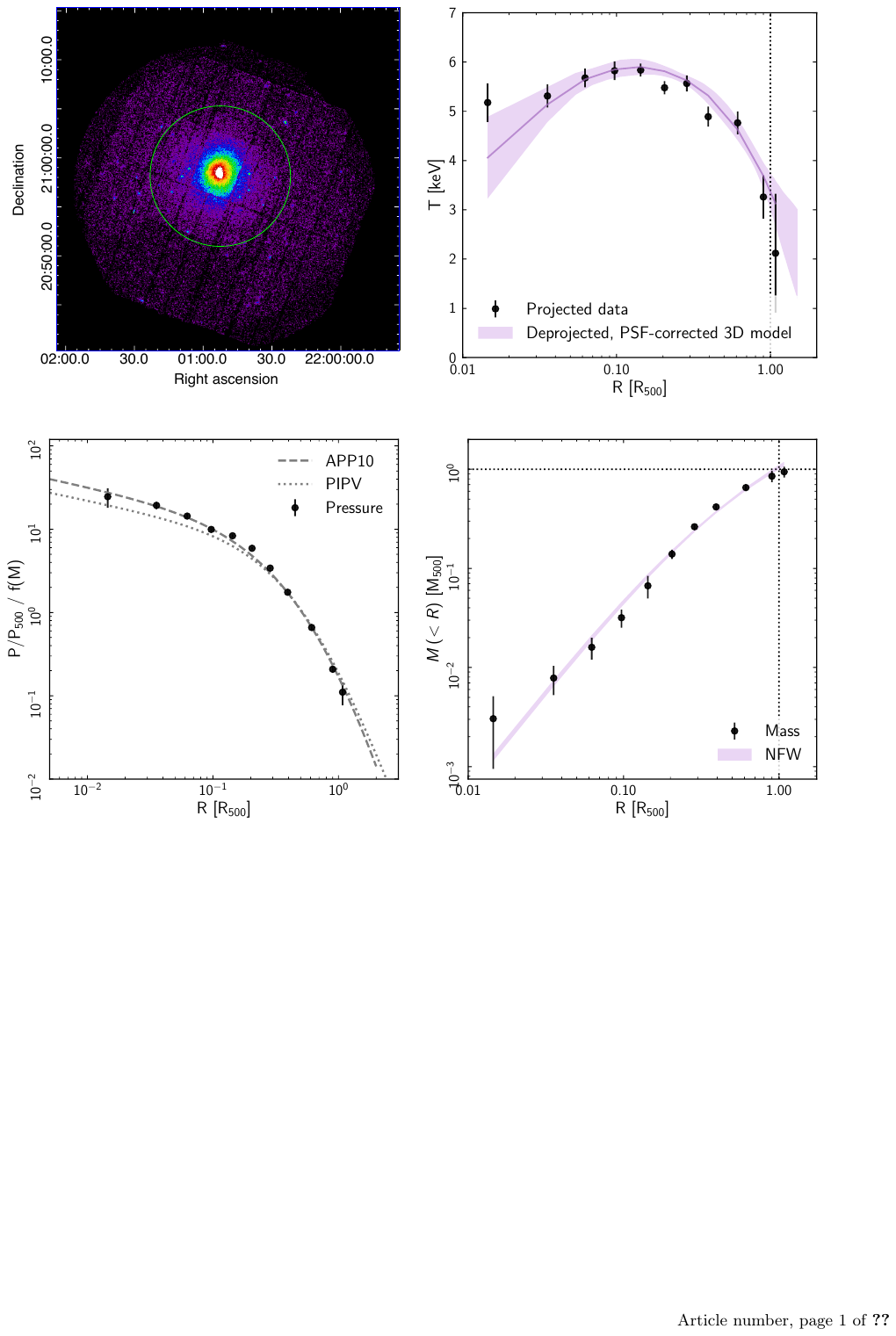}
\caption{\footnotesize Example \xmm\ image and radial profiles of the newly-observed cluster PSZ2\, G$077.90-26.63$, detected at an S/N$\sim 178$ in the $[0.3-2]$ keV band in the $[0.15-1]\,\Rv$ region. {\it Top left}: [0.3-2.0] keV image, with green circle indicating $R_{500}$. {\it Top right:} Raw temperature profile (black points) with best-fitting deconvolved, deprojected 3D model and corresponding $1\sigma$ uncertainties (envelope). {\it Bottom left:} Scaled pressure profile compared to the universal pressure profile of \citet[][APP10]{arn10} and the mean stacked profile of \citet[][PIPV]{planck13}. $f(M)$ is a (small) correction for the mass dependence in the pressure profile shape \citep[see][]{arn10}. {\it Bottom right}: Scaled hydrostatic integrated mass profile compared to the best-fitting NFW model with concentration  $c_{500}=2.6^{+0.3}_{-0.2}$.}\label{fig:psz2g077}
\end{figure*}
% --------------

% --------------
\begin{table*}
\centering
\begin{tabular}{lcccccr}
\toprule
\toprule
Category & Number of  &    \multicolumn{3}{c}{Count rate (ct/s)} &   \multicolumn{2}{c}{\xmms\ time  (ksec)} \\
              &  clusters        &min & med   &  max                          &  Clean   & Required\\
\midrule
Observed \& $\Rv<15'$ & 68 & $0.8$ & $2.9$  & $24.0$	& 1850.9 & 782.9 \\
Observed \& $\Rv >15'$ & 3 & $18.2$& $23.3$ & $53.1$	&  420.7 &  0\\
New \& $\Rv<15^\prime$ 	  & 47 & $0.6$ & $2.9$  &  $7.9$	&   0 & 1029.0 \\
\bottomrule
\end{tabular}
\caption{\footnotesize Required exposures. In each category, the count rates (minimum, median and maximum), 
the cleaned \xmm\ PN archived time,
and the requested exposure time (with no overheads), are given.
None of the three targets with $R_{500} > 15'$ require new exposures.
} 
\label{tab:ksec}
\end{table*}
% --------------

% --------------
% M-z, R-CR plots
\begin{figure*}[t]
\centering
\includegraphics[bb=50 550 550 785 ,clip,scale=1.,angle=0,keepaspectratio,width=0.90\textwidth]{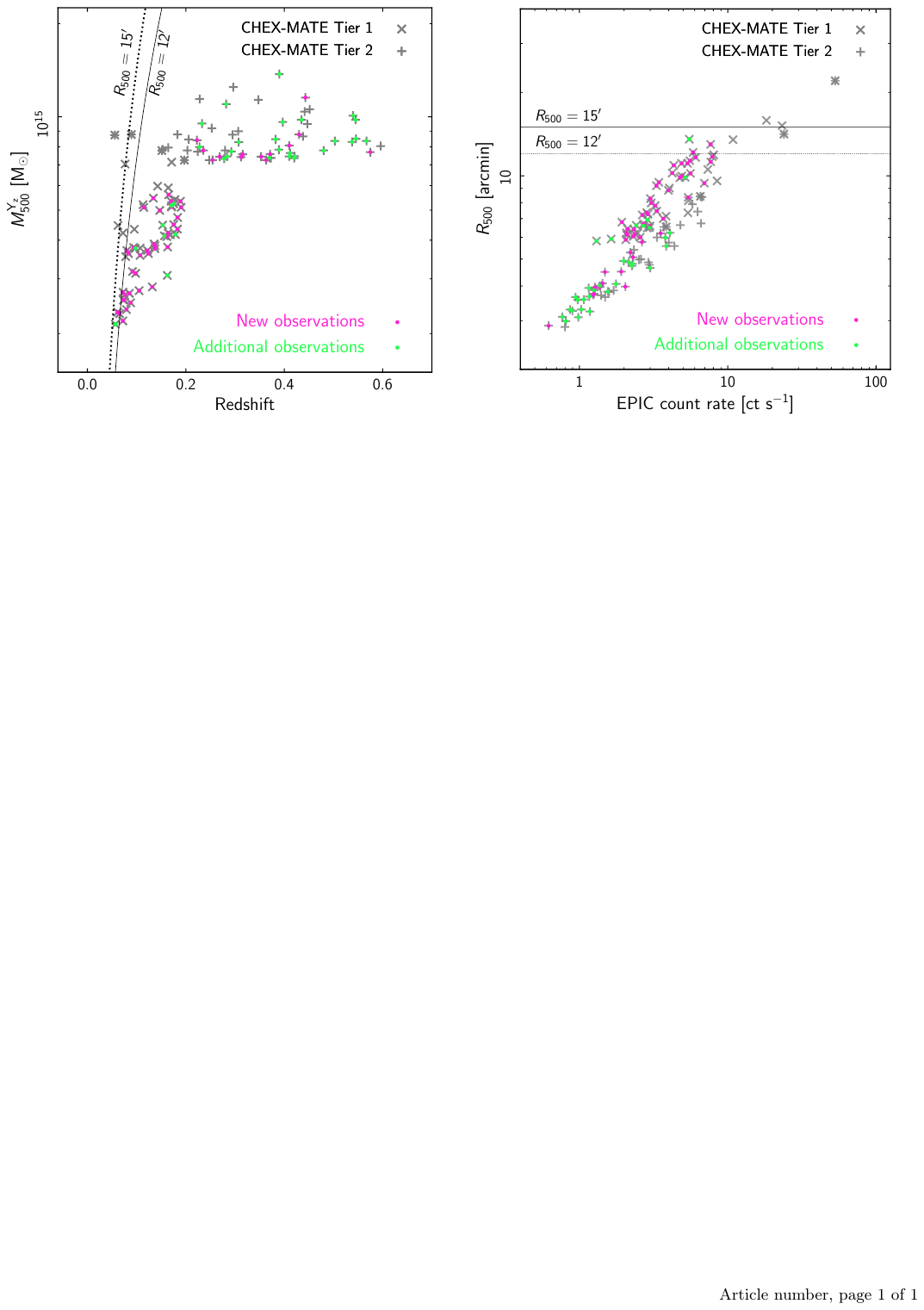}
\caption{\footnotesize {\it Left:} Distribution of the required \xmm\ observations in the $z$--$M$ plane. {\it Right:} Distribution in the size--count rate plane. 
}
\label{fig:zMreq}
\label{fig:CRR500}
\end{figure*}
% --------------

Ultimately, one would like a method to detect clusters based on their most fundamental property: the total mass.  Our project will not be able to  exclude the existence of baryon-poor clusters that are simply not detected in X--ray or SZE surveys. 
Even if we derive the gas properties from X-ray observations, independent of the original SZE detection, there is a residual, intrinsic, covariance with the SZE signal, through the total gas content. Detection of clusters based on their lensing signal, i.e. directly on projected mass, has started to become routinely possible with surveys such as the Hyper SupremeCam Survey \citep[HSC;][]{hsc_miy+al18}.
The {\it Euclid} satellite (and the {\it Rubin} Observatory\footnote{VRO: \url{www.lsst.org/}})  will for the first time allow the detection of sizeable samples of clusters, including the rarest most massive objects, due to their unprecedented sky coverage. Our project has particular synergy with {\it Euclid}, the sensitivity of which should allow blind detection of objects in the redshift and mass range covered by our sample (Fig.~\ref{fig:lens}). Comparison of SZE and shear-selected samples will be critical to assessing residual selection effects, if any.  
It will also be possible to extract high-quality individual and/or stacked shear profiles from {\it Euclid} data, as discussed in more detail in Sect.~\ref{sec:lens}.
The (nearly) all-sky coverage of the Tier-2 sample at high mass will provide the best targets for future strong lensing studies. As the most powerful gravitational telescope in the Universe, they will be high-priority targets for the {\it James Webb} Space Telescope (JWST\footnote{JWST: \url{www.stsci.edu/jwst}}).
In the longer term, our sample will provide the targets of reference for dedicated Athena\footnote{Athena:   \url{www.the-athena-x-ray-observatory.eu/}} 
pointings for deep exploration of ICM physics both in representative (Tier-1) and extreme (Tier-2) clusters.

%%%%%%%%%%%%%%%%%%

\section{Observing strategy}\label{sec:strat}

\subsection{Sample definition}
\label{sec:sample}

The sample is extracted from the \planck\ PSZ2 catalogue \citep{psz2}, including only sources detected in the cosmological mask,  which is the cleanest part of the sky \citep{planck16_sz}. We then excluded the sky region with poor \xmm\ visibility (median visibility less than $55$\, ksec per orbit), which is located in the North (see Fig.~\ref{fig:lens}). We applied a further cut imposing the signal-to-noise ratio (S/N) measured by the MMF3 detection method \citep{melin06} to be larger than $6.5$, allowing us to have a well-controlled analytical selection function. 

This parent sample includes 329 sources, all validated as clusters with $z$ estimates, except for two objects,  PSZ2 G237.41-21.34 and PSZ2 G293.01-65.78. It is a sub-sample of the cosmological sample analysed by \citet{planck16_sz}, but with a slightly higher S/N cut and a more restricted sky region due to the addition of the \xmm\ visibility criteria.  Tier-1 consists of the 61 local $0.05<z<0.2$ clusters in the Northern sky (${\rm Dec} >0$).
In this region, the validation is now  $100\%$ complete \citep[][Dahle et al. in prep.]{bar18_psz1,agu19}, and the overlap with the CFIS survey \citep{cfis} is maximised. The Tier-1 sample has a median mass of $\Mv=4.1\times 10^{14}\,\msol$, as compared to $5.9\times 10^{14}\,\msol$ for the \planck\ Early SZ (ESZ) sample \citep{esz}. 
Tier-2 includes all 61 clusters above $M_{500} > 7.25 \times 10^{14}\,\msol$, as estimated from the MMF3 SZE signal, at  $z<0.6$. 
For this sample of the rarest massive clusters, we had to consider the full parent sample, which at the time of proposal submission was not fully validated. However, the SZE flux of the two sources with missing validation information is such that they would not enter into the Tier-2 selection even if they lie at redshift $z<0.6$. Four clusters are common to Tiers-1 and 2, for a total of 118 clusters, 47 of which have never been observed with \xmm. 

The sample distribution in the $z$-$\Mv$ plane is shown in Fig.~\ref{fig:zM}, and its distribution on the sky is shown in Fig.~\ref{fig:lens}. 
The details of the selection process in the $z$--$M_{500}$ plane is further illustrated in Fig~\ref{fig:appx} in Appendix~\ref{sec:appx}.

% --------------
% Boresight figure
\begin{figure}
\centering
\resizebox{\columnwidth}{!} {
\includegraphics[scale=0.999]{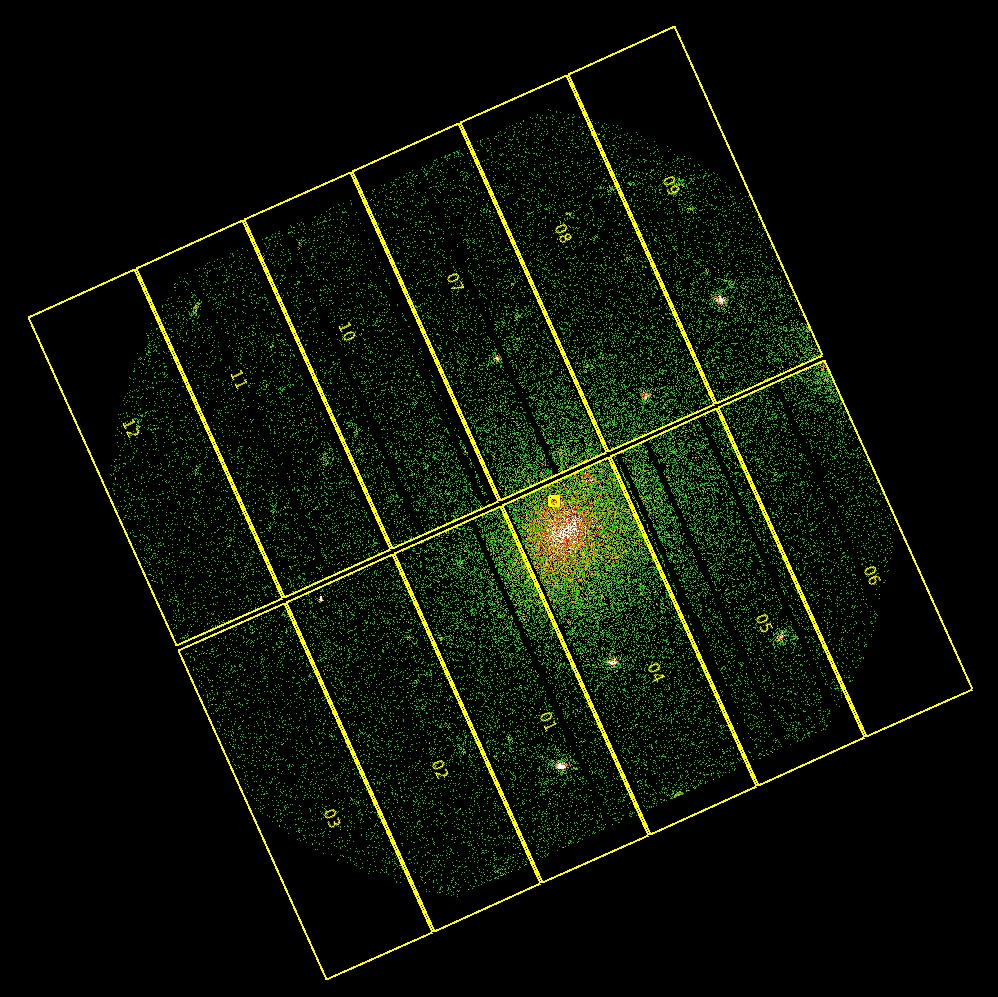}
\hspace{2pt}
\includegraphics[]{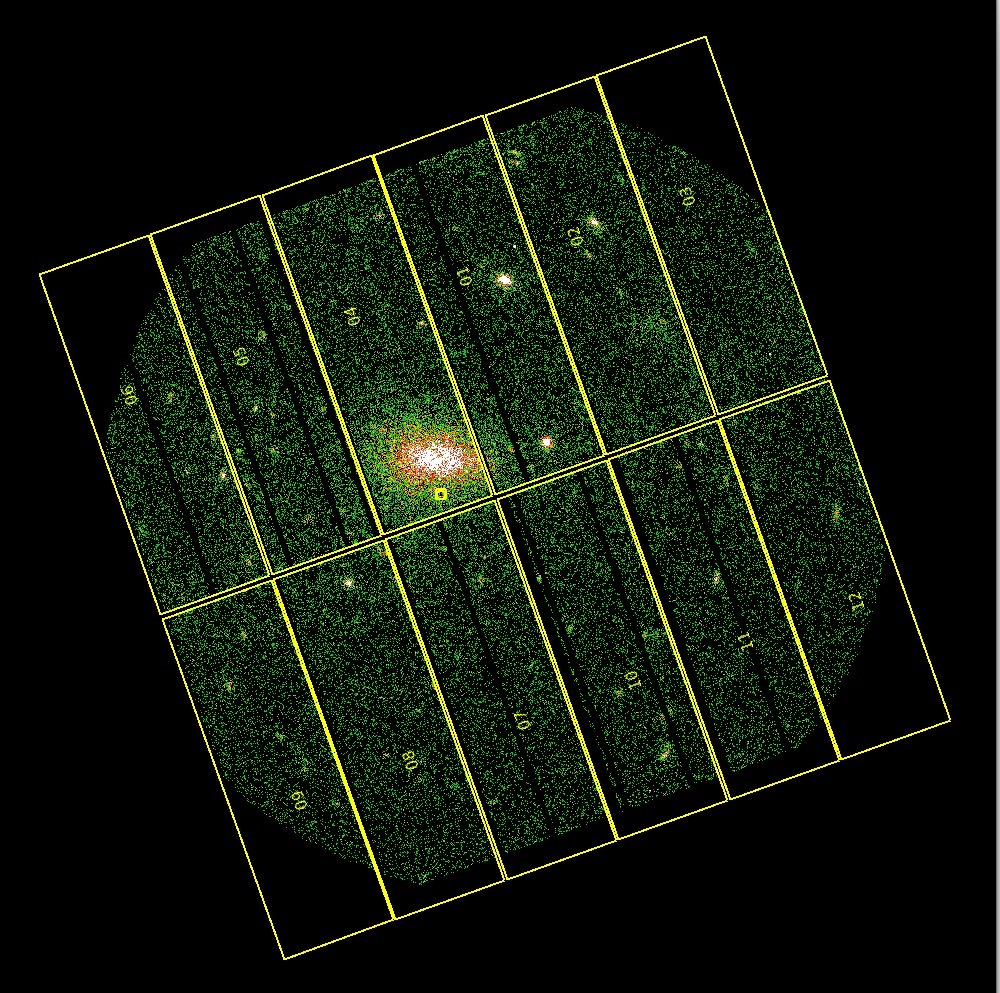}
\hfill
}
\caption{\footnotesize Illustration of the boresight strategy. {\it Left panel:} PN image of PSZ2 G057.78+52.32, at $z=0.0654$, {\it Right panel:} PN image PSZ2 G066.41+27.03,  at $z=0.575$. The yellow grid shows the layout of the PN camera with the nominal boresight marked with a small box.  Depending on the roll angle, the observation boresight is moved $2^\prime$ along pn CCD~4. This avoids the cluster centre region being affected by gaps between CCD chips.  }\label{fig:boresight}
\end{figure}
% --------------

%%%%%%%%%%%%%%%%%%

\subsection{X-ray observation setting}

\subsubsection{Exposure time}

The key observation driver is to obtain temperature profiles up to $\Rv$. We used the mass obtained from the SZE mass proxy, $M_{500}^{\rm YSZ}$,  estimated %iteratively 
from the $\YSZ$ signal \citep{psz1} to obtain the corresponding radii.   
From our analysis of \planck\ clusters \citep{planck2011-5.2b}, we find a tight correlation between $\Mv$ and the core excised luminosity in the soft $[0.5$--$2]$\,\keV\ band when scaled according to purely self-similar evolution, in agreement with the \rexcess\ X-ray sample.  
The expected soft band count rates in the core excised region ($[0.15$--$1]\Rv$) are therefore expected to be particularly robust. 
The conversion between the luminosity and \xmm\ European Photon Imaging Camera (EPIC; PN $+$ MOS) counts takes into account the Galactic column density  ($N_{\rm H}$) value and redshift.   
We checked that the predicted count rates are consistent with those observed for the ESZ-\xmms\ archival sample we have already analysed \citep[see e.g.][]{planck2011-5.2b,lov17} . 
If we define the count rate from the source, the background, and the total as $C_{\rm s} = CR_{\rm s} \times t_{\rm exp}$, $C_{\rm b} = CR_{\rm b} \times t_{\rm exp}$, and $C_{\rm t} = (CR_{\rm s}+CR_{\rm b}) \times t_{\rm exp}$, respectively, then the S/N within the core excised region is, assuming a Gaussian error propagated in quadrature,
\begin{equation}
{\rm S/N} =  \frac{C_{\rm s}}{\sqrt{C_{\rm t} +C_{\rm b}}} =\frac{CR_{\rm s} \sqrt{t_{\rm exp}}}{\sqrt{CR_{\rm s}+2 \times CR_{\rm bkg}}}.
\label{eq:s2n}
\end{equation}
Here, we define the core excised region as $\pi\,(1.-0.15^2)\,\Rv^2$, and adopt $CR_{\rm bkg}\sim1.3 \times 10^{-2}\,{\rm cts\ s^{-1}\ arcmin^{-2}}$ in the [0.3-2]\, keV band.

We set the  exposure time, $t_{\rm exp}$, to reach an S/N$=150$. From our study of ESZ-XMM data, this is sufficient to map the temperature profile in $8+$ annuli at least up to $\Rv$ with a precision of  $\pm15\%$ in the $[0.8$--$1.2]\Rv$ annulus, and to reach an uncertainty of $\pm 2\%$ (statistical uncertainty) on the mass derived from the $Y_{\rm X}$ mass proxy, $\Mv^{YX}$, 
and to derive the HE mass measurements at $\Rv$ to the $\sim15$--$20\%$ precision level.  
The precision is illustrated in Fig~\ref{fig:psz2g077}, where we show an analysis of the representative observation of PSZ2\,G$077.90-26.63$, which reaches the required S/N. 

With regard to archival \xmm\ observations, we processed all archival observations (including offset pointings) of Tier-1 and Tier-2 ($71$ clusters in total) to estimate the clean (soft proton flare-free) time of the PN camera. This was subtracted  from the requested time. Thirty-three clusters needed re-observations. They are marked in Fig.~\ref{fig:zMreq} with green points, together with the 47 clusters that have never been never observed with \xmm\ before (pink points).  

The $\Rv$ size of three clusters is larger than  the \xmm\ $15'$ field of view (see Fig.~\ref{fig:CRR500} and Table~\ref{tab:ksec}); for these three objects, offset pointings are available in the archive to enable detection of the ICM to $\Rv$. For four clusters with $12'<\Rv<15'$ and without offset observations in the \xmm\  archive, we required one extra $15$\, ksec pointing for precise background measurements.

The final total project observing  time is summarised in Table~\ref{tab:ksec}. The required time was increased by $40\%$  to account for time loss owing to soft proton flares, and a minimum exposure time of 15 ksec was set to enable efficient use of \xmm\ (in view of observation overheads and slew time). The final list of CHEX-MATE target observations, including archival observations, is presented in Tables~\ref{tab:master1b}, \ref{tab:master2b}, and \ref{tab:master3b}. These tables list all target properties that were used in the selection and exposure time estimation.

% --------------
% XMM image gallery
%
\begin{figure*}[!ht]
\centering
\includegraphics[width=0.95\textwidth]{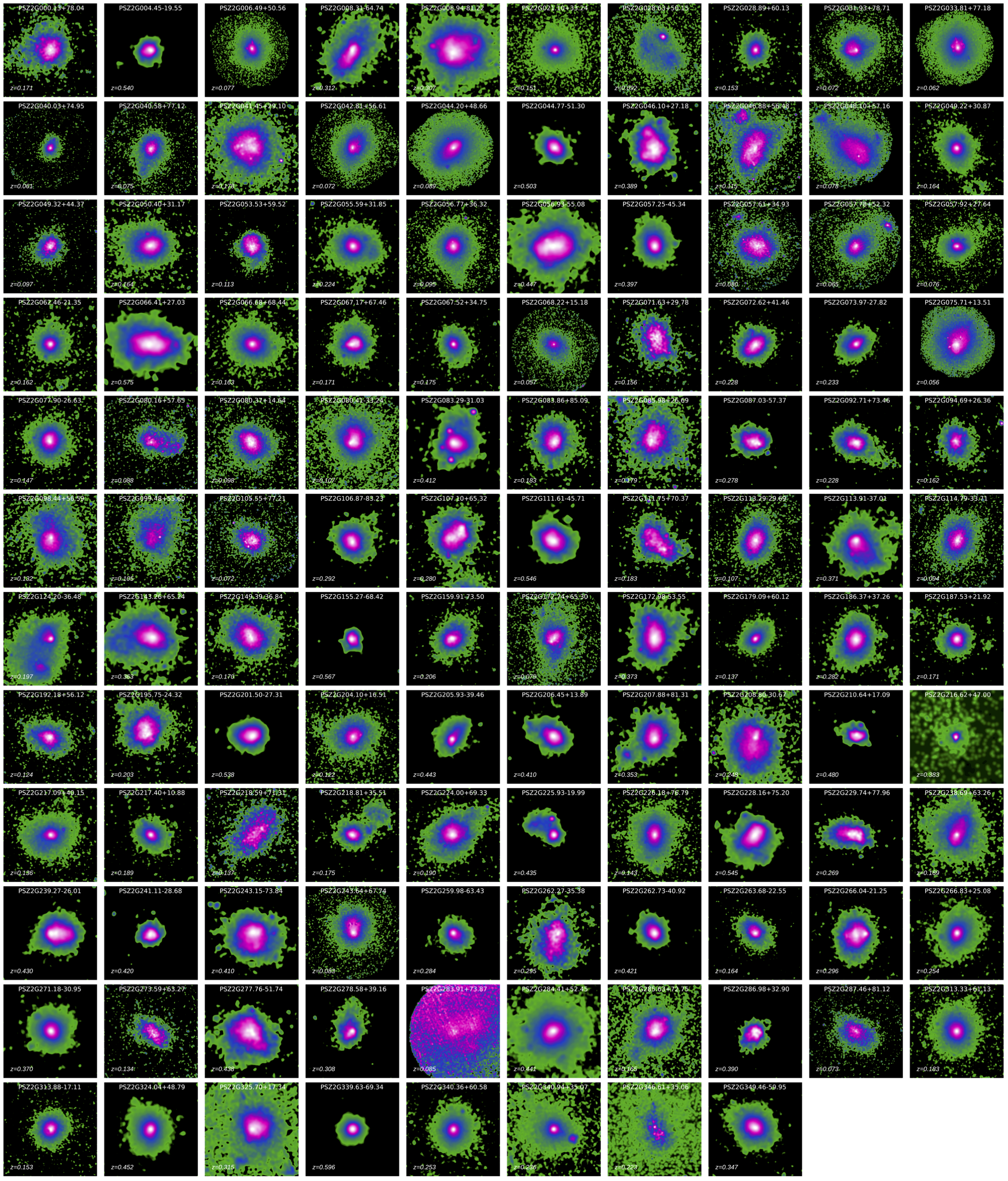}
\caption{\footnotesize 
\xmm\ image gallery of the 118 targets.
The images cover an area of 2.4 $R_{500} \times$ 2.4 $R_{500}$. After the main point sources have been masked and their emission has been replaced with an average contribution from the nearby environment, they are background subtracted, exposure corrected and smoothed with a Gaussian with $\sigma$ = 7.5 arcsec. Low-quality images correspond to objects for which the exposure time will be completed in the final year of observations. 
}
\label{fig:gallery}
\end{figure*}
% --------------
% --------------
\begin{table}[]
    \centering
\begin{tabular}{@{}lcccccr@{}}
\toprule
\toprule
     Name & RA    & Dec  & obsID & $t_{\rm clean}$ \\
      & h:m:s & d:m:s &   & ksec \\
\midrule
PSZ2 G113.91-37.01 & 00:19:39 & +25:17:27 & 21059 & 9.8 \\
PSZ2 G217.40+10.88 & 07:38:19 & +01:02:07 & 21060 & 9.6 \\
PSZ2 G083.86+85.09 & 13:05:51 & +30:54:17 & 21061 & 9.8 \\
PSZ2 G149.39-36.84 & 02:21:34 & +21:21:57 & 21062 & 9.9 \\
PSZ2 G062.46-21.35 & 21:04:54 & +14:01:40 & 21063 & 9.9 \\
PSZ2 G218.59+71.31 & 11:29:55 & +23:48:14 & 21064 & 9.8 \\
PSZ2 G028.63+50.15 & 15:40:09 & +17:52:41 & 21065 & 9.8 \\
PSZ2 G080.16+57.65 & 15:01:08 & +47:16:35 & 21066 & 10.3 \\

\bottomrule
    \end{tabular}
    \caption{\footnotesize List of the eight  objects with a dedicated \cxo\ snapshot. RA and Dec refer to the \cxo\ pointing coordinates; obsID identifies the \cxo\ Observation identification number; $t_{\rm clean}$ indicates the cleaned exposure time.}
    \label{tab:ddt}
\end{table}
% --------------

%%%%%%%%%%%%%%%%%%

\subsubsection{Cluster centre and pointing position}

We optimised the position of the cluster cores in the \xmm\ field-of-view to avoid the PN camera CCD chip gaps crossing the central region of the object. This was achieved by moving the centre from the nominal boresight position by 2$’$ away from the gap, along the PN CCD 4. This strategy is illustrated in Fig.~\ref{fig:boresight}, which shows the new observation of the nearby Tier-1 cluster PSZ2\,G$057.78+52.32$, at $z=0.0654$, and the distant Tier-2 cluster PSZ2\,G$066.41+27.03$, at $z=0.575$. 

The  new boresight depends both on the cluster position and the position (roll) angle of the observation, which is not known in advance of scheduling. We thus computed a grid of boresight values  versus roll angle. For some specific clusters with interesting sub-structure, the position was further refined (only for the possible angle of the orbits where the cluster is visible). We very much benefited from the help of the \xmm\ SOC for project enhancement in this procedure, who implemented the optimised boresights for each observation.

This strategy requires a good knowledge of the position of the cluster centre. The uncertainty on the Planck position, which is $2^\prime$ on average and can reach $5^\prime$, is too large for our  purpose \citep{psz1}.  We relied on X--ray positions retrieved from archival data for 72 clusters. This includes  the 33 clusters with previous \xmm\  observations, 32 clusters with \cxo\ data, and seven  clusters with sufficiently deep Swift-XRT observations and/or ROSAT observations. 

The CHEX-MATE sample includes eight clusters (PSZ2 G028.63+50.15, PSZ2 G062.46-21.35, PSZ2 G080.16+57.65, PSZ2 G083.86+85.09, PSZ2 G113.91-37.01, PSZ2 G149.39-36.84, PSZ2 G217.40+10.88, PSZ2 G218.59+71.31) that have no \cxo, \xmm, Swift-XRT, or RASS exposures.
We obtained snapshot \cxo\ observations of about 10 ksec duration for each of these systems through joint \cxo-\xmm\ time (see Table~\ref{tab:ddt}). These data allow us to detect the emission of each object, to confirm the X-ray centres, and to obtain a preliminary indication of the X-ray morphology. 

%%%%%%%%%%%%%%%%%%

\subsection{X-ray data quality assessment and analysis procedures}

\xmm\ began observing the sample in mid-2018, and the observation programme will last three years. We reduce new observations as soon as they become available in the \xmm\ archive to assess their quality by computing several indicators: the fraction of clean time (after removal of soft proton flares) with respect to $t_{\rm exp}$ estimated from Eqn.~\ref{eq:s2n}, the S/N, and the count-rate in the core-excised region. We also compute the level of particle background induced by galactic cosmic rays (as measured by the count rate in the detector region outside the MOS field of view) and the level of residual contamination in the field of view \citep[see e.g.][]{cxb, sal17}. We also perform a full standard analysis up to the production of the hydrostatic mass profile. 

At the end of the second year of observations, we used this information to decide whether some of our targets would require a re-observation to reach our objective during the third and final year of observations. We found 15 observations for which the S/N in the core excised region was smaller than 90\% of our goal (Eq.~\ref{eq:s2n}), and we looked at the complete analysis to prioritise them. We also noticed that one of the offset observations we requested and the observations of two clusters of our sample performed in AO17 under different programs were badly affected by soft proton flares. 

We were able to accommodate re-observation of ten targets within our time budget by reducing the overheads of each observation in the last year. We changed the observation mode from Extended Full Frame to Full Frame, and withdrew the observations of four clusters (PSZ2\,G092.71$+$73.46,  PSZ2\,G049.32$+$44.37, PSZ2\,G073.97$-$27.82, PSZ2 G073.97-27.82) for which the exposure time of archival observations was already larger than $0.8\,t_{\rm exp}$ after checking the quality of their temperature and mass profile. 

\xmm\ observations of the full sample will be  reduced and analysed by combining the best practices developed during previous projects, such as  
\rexcess\ \citep[][]{cro08,pra09,pra10}, \planck\ \citep[][]{pipIII,planck13}, X-COP \citep{tch16,ghi18,eck19}, and M2C \citep[][]{bartal18,bartal19}. The final pipeline will emphasise the complementarity of the methods developed in these projects (e.g., point spread function correction, accounting for gas clumping), and we are also developing new and innovative techniques within the collaboration.  We will use \xmm\ photons in an energy band that maximises the source-to-background ratio to derive surface brightness and density profiles up to $\Rv$ and beyond, and to measure quantitative morphological indicators within $\Rv$. We will apply a full spectral modelling of the \xmm\ background to measure radial profiles with a statistical uncertainty of $15\%$ on the temperature estimate at $\Rv$, from which we will derive high-accuracy profiles of thermodynamic quantities and total mass, with both parametric and non-parametric methods \citep{cro06,dem10,ett10,ghi18,ett19,bartal18}. Statistical properties for the full sample, such as mean profiles, scaling laws, and the scatter around them, will be derived in self-consistent way \citep[e.g.][]{mau14,ser16_lira}. 
The details of the data analysis will be discussed in forthcoming papers. Final data will be made available in a dedicated public database of integrated quantities and reconstructed profiles. 

A preliminary gallery of the smoothed X-ray surface brightness maps is shown in Fig.~\ref{fig:gallery}. The images have been exposure corrected, background subtracted, and point sources have been removed and replaced by an average contribution from the nearby environment.

%------------
% Lensing Figure 1 - z-M
%
\begin{figure}[t]
\centering
\includegraphics[width=0.475\textwidth]{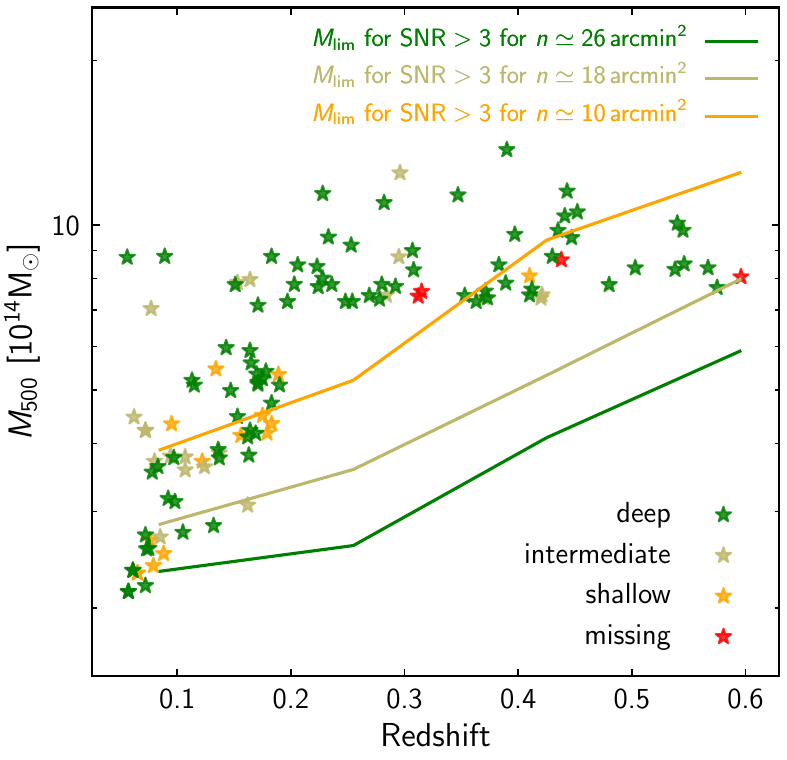}
\caption{\footnotesize 
Lensing observational status. $\Mv$--$z$ plane with symbol colours representing the depth of the archival or dedicated lensing data. Shallow surveys like CFIS (and more so, DES, KiDS) yielding a density of background sources $n\lesssim 10\, {\rm arcmin}^{-2}$ cannot probe the low mass end of Tier-1 clusters with ${\rm S/N} > 3$. Stacking of the shear signal will be unavoidable for these. Deep Subaru data enable such measurements on individual clusters since most observations reach source densities $\gtrsim 20\, {\rm arcmin}^{-2}$. {\it Euclid}, which will reach $n\simeq 30\, {\rm arcmin}^{-2}$ will greatly simplify cluster mass calibrations with lensing.  
}
\label{fig:lens1}
\end{figure}
%------------

%%%%%%%%%%%%%%%%%%

\section{Supporting data}
\label{sec:support}

\subsection{Lensing}
\label{sec:lens}

Accurate WL measurements of the matter distribution of the CHEX-MATE clusters are crucial to fulfilling the project goals. The homogeneous and complete WL coverage of the sample can be obtained by complementing high-quality optical archival data from ground based telescopes with dedicated proposals.

%------------
% Lensing Figure 2 - PSZ G077.9
%
\begin{figure*}[t]
\centering
\includegraphics[bb=50 580 570 785 ,clip,scale=1.,angle=0,keepaspectratio,width=0.975\textwidth]{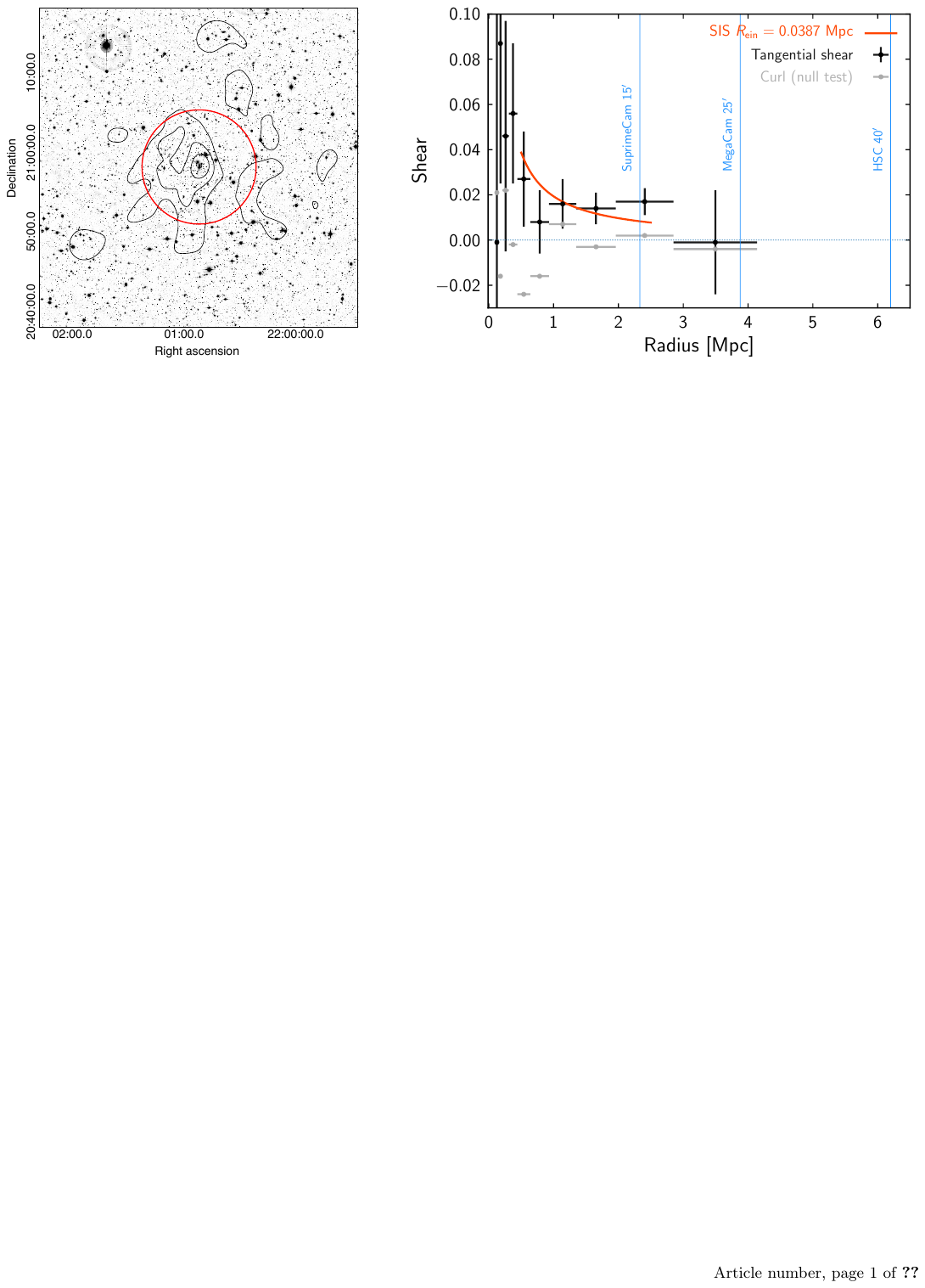}
\caption{\footnotesize 
{\it Left:} MegaCam $r$-band optical image of the Tier-1 cluster PSZ2\,G077.90-26.63 (A2409) of mass $M^{Y_{\rm SZ}}_{500} \simeq 5\times 10^{14} \msol$ at redshift $z=0.148$, with convergence (i.e. surface mass density) contours overlaid. The red circle delimits $\Rv$. {\it Right:} Radial shear profile obtained for PSZ2 G077.90-26.63, showing that, besides the ability to constrain the overall mass to within $30\%$ or so, only a modest handle on individual profile shapes is possible. The vertical lines show the outermost scale grasped by the main facilities we use for lensing, which nicely encompass the virial radius of most clusters, even at low redshift. In particular, the $0.5-2.5\, {\rm Mpc}$ range (radial span of the red curve) over which model fitting performs best, is well covered by our data. Grey points represent the curl component of the shear profile, which should be consistent with zero (error bars are not shown for clarity, but are the same as for the tangential component).
}
\label{fig:lens2}
\end{figure*}
%------------

More than half of the sample, 62 clusters out of 118, have already been studied as part of various published WL analyses; these are detailed in the Literature Catalogs of weak Lensing Clusters of galaxies \citep[LC$^2$;][]{ser15_comalit_III}\footnote{LC$^2$ are standardised meta-catalogues of clusters with measured WL mass, available at \url{pico.oabo.inaf.it/\textasciitilde sereno/CoMaLit/LC2/}.}. Programmes including CLASH \citep{ume+al14}, WtG \citep[][]{wtg_III_14}, CCCP  \citep[][]{her+al19}, LoCuSS \citep[][]{ok+sm16}, or PSZ2LenS \citep{ser+al17_psz2lens} have shown that WL analyses can recover the mass up to a best accuracy of $\sim 20-25\%$ (including scatter due to triaxiality, substructures, intrinsic shape, and cosmic noise;~e.g.\citealt{ume+al16b}).

For lensing, the best possible multi-band optical wide field imaging is required. We thus consider observations with the 8.2-m Subaru telescope with the Hyper Suprime Cam (HSC) ($1.77~\deg^2$ FoV)\footnote{HSC: \url{www.naoj.hawaii.edu/Observing/Instruments/HSC/index.html}} and its SuprimeCam ($34\arcmin\times 27\arcmin$ FoV) precursor \citep{hsc_miy+al18,hsc_kom+al18,hsc_fur+al18,hsc_kaw+al18,miy+al02a} along with MegaCam at the 3.6-m Canada-France-Hawaii Telescope (CFHT) ($1 \deg^2$ FoV) \footnote{MegaCam: \url{www.cfht.hawaii.edu/Instruments/Imaging/MegaPrime/}}, both located at the Mauna Kea summit (Hawaii). For the Southern hemisphere, the OmegaCam\footnote{Omegacam: \url{www.eso.org/sci/facilities/paranal/instruments/omegacam.html}} at the 2.5-m VLT Survey Telescope (VST) on Paranal (Chile) ($1 \deg^2$ FoV) and the Wide Field Imager (WFI)\footnote{WFI: \url{www.eso.org/sci/facilities/lasilla/instruments/wfi.html}} at the 2.2-m MPG/ESO ($0.25 \deg^2$ FoV) telescope at La Silla (Chile) are also considered. Good partial or complete data sets are already available from these archives for 83 clusters.

Additionally, two ongoing surveys are of particular interest for the CHEX-MATE program. The Hyper Suprime-Cam Subaru Strategic Program \citep[HSC-SSP,][]{hsc_miy+al18,hsc_aih+al18} has been carrying out a multi-band imaging survey in five optical bands ($grizy$) with a depth of $i \sim26$ at the $5\sigma$ limit within a 2\arcsec\ diameter aperture, aimed at observing $\sim1400\deg^2$ on the sky in its Wide layer \citep{hsc_aih+al18}. The survey is optimised for WL studies \citep{hsc_man+al18,hsc_hik+al19,hsc_miy+al19,hsc_ham+al19,oka19} and should be completed in 2021. Five CHEX-MATE clusters fall in the HSC-SPP footprint.
CFIS is an ongoing legacy programme at the CFHT \citep{cfis}. It is part of a wider multi-band imaging effort named UNIONS, which is underway to map the Northern extragalactic sky, notably to support the {\it Euclid} space mission. To aid the follow-up of CHEX-MATE, 33 Tier-1 clusters have expressly been selected to lie in the CFIS footprint. About $\sim4500\,\deg^2$ will be obtained in the r-band to a depth of 24.1 (point source, S/N=10, 2\arcsec\ diameter aperture) with a median seeing of $0\farcs66$. As of now, $2500\,\deg^2$ are already available and full completion may require another two years of observations. 
CFIS observations in the u-band  (mag$_{\rm lim}\sim 23.6$, median seeing $0\farcs85$), are not deep enough to bring significant photometric information for the background sources but will aid our understanding of the star formation in cluster member galaxies. Likewise, complementary z band data coverage in the UNIONS collaboration is being obtained with good image quality from Subaru (WISHES program, PI M. Oguri), which has also started to observe the same footprint to a magnitude of 23.4 (same definition as above). This is comparable to the r-band depth, and can thus be helpful for the stellar mass content of cluster member galaxies as well as for the redshift estimation of the faint background sources.
In total, 34 clusters (nine unique clusters covered neither by archival data nor dedicated proposals) fall in these two survey footprints.

The data set will be completed with targeted observations of 31 clusters from dedicated proposals (26 unique clusters not covered at all by archival data) or ongoing WL surveys for 34 clusters (nine unique clusters). The CHEX-MATE collaboration has already been awarded $\sim 32$~h at HSC@Subaru (proposals S19B-TE220-K, S20A-TE129-KQ, S20B-TE212-KQ, P.I. J.~Sayers), $\sim21$~h at Megacam@CFHT (P.I. R.~Gavazzi/K.~Umetsu), and $\sim23$~h at OmegaCam@VST (proposals 0104.A-0255(A) and 105.2095.001, P.I. M.~Sereno). 
A partial summary of already available observations is reported in Table~\ref{tab_lensing_data}. Some redundancy in available data is present and this will be exploited to assert our control of systematics in shear measurements by requiring consistency between lensing data measured with HSC and Megacam for instance. A full assessment of the quality and internal consistency of the lensing measurement will be addressed in specific papers.

Arguably, the driving criterion for obtaining accurate lensing measurements is the surface number density of background, potentially lensed, sources, lying far behind the foreground massive cluster.
With observations with integration time of $\sim$ 30 minutes at an 8-metre class telescope, source densities as high as $n_{\rm bg} \simeq 20\, {\rm arcmin}^{-2}$ \citep[eg][]{hsc_med+al18b} can be obtained. Hence, the lensing signal from regions up to $\sim 2-3~\text{Mpc}$ can be recovered with an S/N$\sim5-10$ \citep{wtg_III_14,ok+sm16,ume+al16b}. For comparison, {\it Euclid} space-borne imaging should routinely yield densities $n_{\rm bg}\simeq 30\, {\rm arcmin}^{-2} $.
With CFHT and similar telescopes, reaching the same depth is more difficult and most often lensing data deliver $n_{\rm bg}\sim9-15\, {\rm arcmin}^{-2}$, (with a 30-60 minute integration time). This is particularly true for CFIS. Shallower surveys like KiDS or DES do not exceed $n_{\rm bg} \sim8\, {\rm arcmin}^{-2}$. The point spread function represents an additional problem for ground-based observations, as an increase in the number of blended sources reduces the number of galaxies that can be used for WL. As shown in Fig.~\ref{fig:lens1}, deep observations corresponding to the best images ($t>30 $ min on Subaru) and observations of intermediate depth ($30>t>10$ min on Subaru-equivalent telescopes\footnote{For other telescopes, the equivalent exposure time is rescaled by the square of the primary dish diameters to account for differences in telescope sensitivity levels.}) should enable individual mass measurements of 33\% accuracy or better for most Tier-1 and all Tier-2 clusters. The shallower data ($t<10$ min on Subaru-equivalent) will not permit such mass determinations on individual clusters, so one would have to resort to stacking techniques in order to put constraints on the lowest mass end ($M_{500}\lesssim 3\times 10^{14}\,\msol$) of Tier-1 clusters.

Depth is not the only criterion, however. Some amount of colour information on background sources is required for efficient and clean separation of background galaxies from cluster members and foreground sources. A two-band colour selection is needed for clusters at $z \la 0.2$, whereas three bands are needed for more distant clusters. With this requirement, we are able to control contamination by cluster member galaxies at the percent level \citep{ok+sm16} and, whenever needed, our dedicated observations will obtain this minimal coverage. For many of the well-known Tier-2 clusters, several more bands are often available (uBV, Rc, I, z), and will be used. 

%------------
% SZ figure 1 : z-M coverage
%
\begin{figure}[t]
\centering
\includegraphics[width=0.5\textwidth]{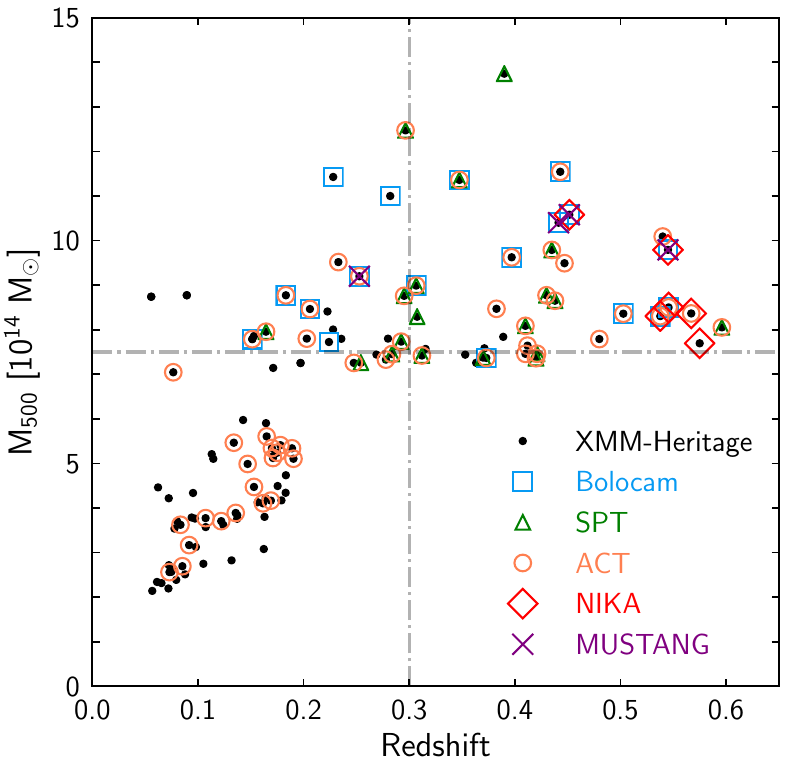}
\caption{\footnotesize 
Current SZE coverage by various facilities of the CHEX-MATE sample in the $M_{500}-z$ plane. The details are given in Table~\ref{tab:sz} and are further discussed in  Sect.~\ref{sec:sz}. 
}
\label{fig:sz}
\end{figure}
%------------

In addition to an overall mass measurement, WL can also provide information on the mass density profile if the density of background sources is large enough ($n \gtrsim  20\, {\rm arcmin}^{-2}$). The right-hand panel of Fig.~\ref{fig:lens2} shows the radial shear profile one can obtain under the typical observing conditions we expect. The example is PSZ2\, G$077.90-26.63$ (A2409) at $z=0.148$, for which deep SuprimeCam data yields $n=22$ faint background galaxies per square arc minute out to about 2.5 Mpc from the centre. The accuracy on mass is 33\% for a mass of order $M_{500} \simeq 5\times 10^{14} \msol$. 
We typically expect the shear signal to deliver constraints on the concentration of individual halos to 30\% accuracy for the most massive clusters, with a source density $n \gtrsim  20\, {\rm arcmin}^{-2}$. On the other hand, for low-mass, Tier-1 clusters, with the shallowest (CFIS or DES-like) observations, the same accuracy can only be achieved after the stacking of about 20 clusters or so. In this process, we intend to stack the likelihood in a hierarchical Bayesian manner \citep[see eg][]{lieu17} rather than use a crude shear stacking in concentric annuli.

%%%%%%%%%%%%%%%%%%

%------------
% SZ Figure 2 - PSZ2G 077.9
%
\begin{figure*}[t]
\centering
\includegraphics[bb=50 580 570 785 ,clip,scale=1.,angle=0,keepaspectratio,width=0.975\textwidth]{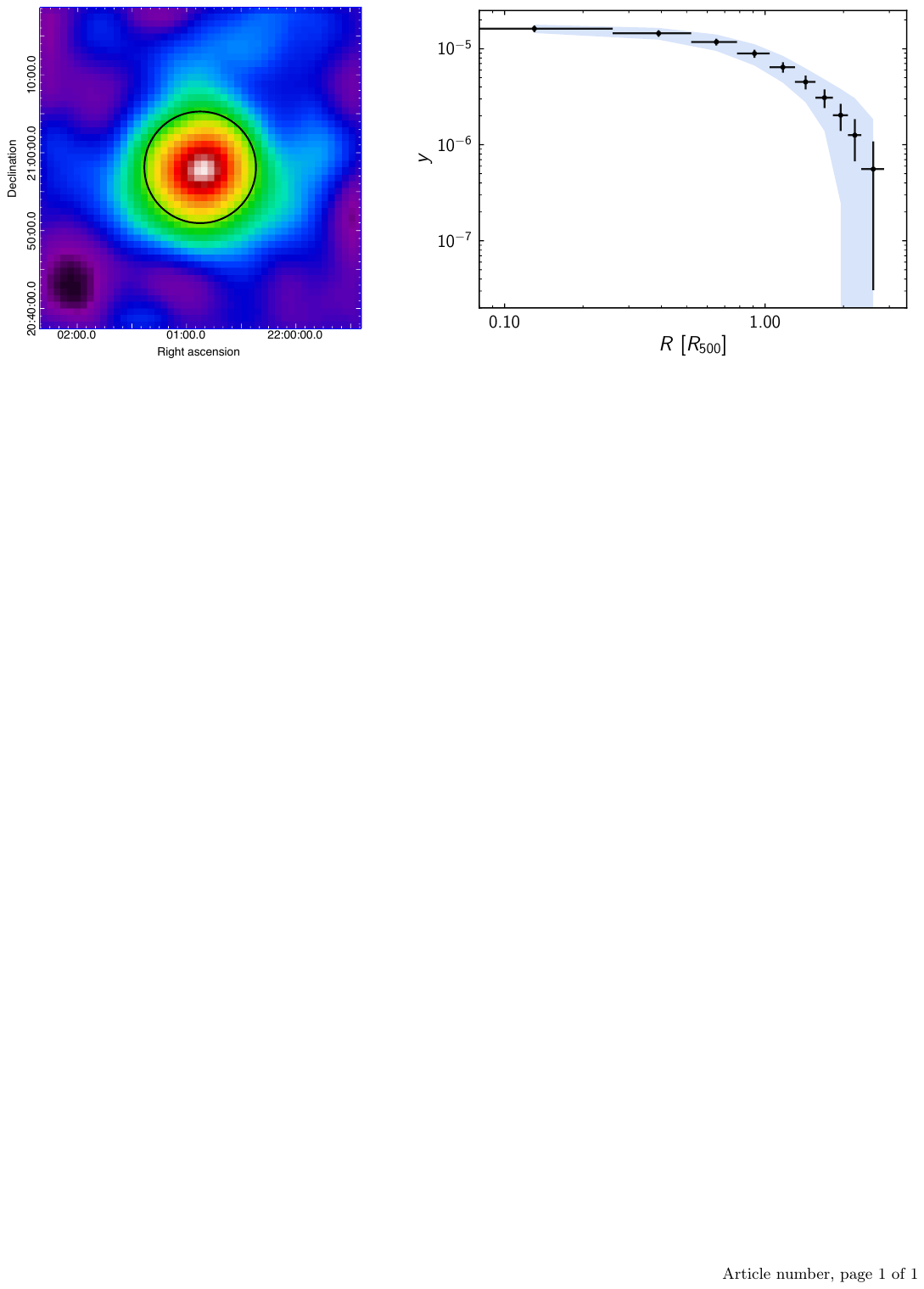}
\caption{\footnotesize 
{\it Left:} \planck\ Comptonisation $y$ map of the Tier-1 cluster PSZ2\,G077.90-26.63. The black circle represents $\Rv$. {\it Right:} Radial SZE profile in Comptonisation parameter, $y$, and radial units of $R_{500}$. The profile is derived from the $y$-map following the method published  in \citet{planck13}. The individual points are correlated at about the 20\% level and the error bars show the square root of the diagonal elements from the covariance matrix. The blue shaded envelope shows the dispersion in flux about the profile.
}
\label{fig:SZ2}
\end{figure*}
%------------

\subsection{Sunyaev-Zeldovich effect}
\label{sec:sz}

As stressed above, the SZE data are complementary to the X-ray data, providing an independent tracer of the hot intra-cluster gas. 
% Planck data 
Our sample of 118 clusters was selected from the {\it Planck} all-sky survey \citep{planck2014-a01} with a S/N $> 6.5$. We therefore have high-quality {\it Planck} SZE data for all of the targets. For example, from the public {\it Planck} all-sky Modified Internal Linear Combination Algorithm (MILCA) SZE map \citep{PlanckDR15}, we can obtain the radial distribution of the SZE signal for each object in our sample. From further deprojection and deconvolution, we can also reconstruct the underlying 3D gas pressure profile following the methodology developed by \citet{planck2012-V}. In conjunction with the \xmm\ data, these {\it Planck}-derived constraints will provide further insights into the scaling and structural properties of the galaxy cluster population.

% Bolocam data 
For the 61 Tier-1 clusters at $z < 0.2$, the {\it Planck} data alone are likely to be sufficient for most desired analyses. For the higher-$z$ Tier-2 clusters, many potential analyses will benefit from the inclusion of higher angular resolution SZE data from wide-field ground-based facilities \citep[see, e.g.][]{say16, rup18}. In particular, data are publicly available from Bolocam \citep[]{Sayers2013_pressure}, the SPT-SZ survey \citep[]{cho18}, and the ACT surveys \citep[]{Aiola2020}. In total, these data include 43 unique Tier-2 clusters (and 21 unique Tier-1 clusters), some with coverage from more than one data set. In the relatively near future, we also expect data releases from the SPT-ECS survey \citep[]{ble20} and the New Iram Kids Array (NIKA2) SZ Large Program \citep[]{Mayet2020}. In total, these data will include five additional unique Tier-2 clusters. A summary of the available SZE data is given in Figure~\ref{fig:sz} and Table~\ref{tab:sz}.

% And the rest
Beyond these wide-field SZE data, which generally have an angular resolution of $\sim 1$~arcminute, ground-based SZE observations with spatial resolution comparable to the X-ray data could provide a transformational added value. Joint X-ray and SZE analyses would allow detailed reconstructions of the internal structure of the physical properties of the hot gas \citep[e.g.][]{ada17,rup18}. In particular, NIKA2 \citep{per20} and MUSTANG-2 \citep{dic14}, currently operating on the Institut de Radioastronomie Millimétrique (IRAM) 30m and Green Bank Telescope (GBT) 100m telescopes, obtain 18 and 9~arcsec FWHM resolutions at 150 and 90~GHz, respectively.

Even higher resolution SZE observations are possible with current large interferometric observatories such as the Atacama Large Millimetre Array (ALMA) and the Northern Extended Millimeter Array) NOEMA \citep[see, e.g.][]{kit16}. Accounting for the limited coverage provided by these facilities, such observations would target, within a reasonable exposure time, specific regions for either a single cluster or a sample of targets; for example, a follow-up of shocks or any other spatial feature of interest \citep{bas16,kit20}.

% objectives 
From the combination of the SZE data from {\it Planck}, along with publicly available data from Bolocam, SPT, ACT, and NIKA2, we will derive global SZE properties such as $Y_{SZ} = \int y d\Omega$, where the Compton $y$ parameter is integrated over the aperture $\Omega$ obtained from the X-ray \xmm\ analysis (centroid, $R_{500}$, etc.) to construct scaling relations (e.g. $Y_X-Y_{SZ}$) for the entire sample. Revisiting previous works \citep[e.g.][]{planck2011-5.2a,planck2011-5.2b}, this will provide a solid local reference from an SZE selected sample covering the full mass range (Tier-1) and and from a mass-limited sample at low-to-intermediate redshift (Tier-2). In addition, joint X-ray and SZE analyses, building on what was performed for the X-COP and CLASH projects \citep{eck17_xcop, sie18,Sereno2018}, will provide a complementary view to standalone X-ray analyses of the structural thermodynamical properties beyond $R_{500}$ and into the clusters' outskirts. High-resolution SZE images will be also instrumental in constraining the ICM power spectrum  jointly with X-ray images \citep[e.g.][]{kha16}. An example image and radial profile of PSZ2\,G077.9-26.63, obtained from the \planck\ survey data, is shown in Fig.~\ref{fig:SZ2}. 

% Follow-up
We would ideally like to obtain SZE data with an angular resolution comparable to the \xmm\ X-ray images. The complementarity of these multi-probe data would allow for detailed studies of sub-structures within the ICM (see e.g. the recent combination of \xmm\ and NIKA2 or MUSTANG data by \citealt{rup18}, \citealt{oka19}, and \citealt{ker20}).
As noted above, both NIKA2 and MUSTANG-2 can provide such data and are available for open-time observations.
For both instruments the integration time goes from reasonable (a few hours) to relatively time consuming ($\sim$10-20h per targets) depending on the mass and redshift of the cluster. For example, NIKA2 could provide images extending to $R_{500}$ of clusters at $z=0.3$ in approximately three hours for $M_{500}=15\times 10^{14}$~M$_\odot$ and in approximately 18 hours for $M_{500}=7\times 10^{14}$~M$_\odot$.
Based on realistic open-time requests, and actual allocations to other large cluster programs \citep[e.g.][]{Mayet2020,Dicker2020}, obtaining coverage for sub-samples of $\lesssim 10$ clusters is possible. We will thus pursue MUSTANG-2 and NIKA2 imaging of well-defined sub-samples, or individual targets, where the high angular resolution SZE data will have the most impact. In addition, such followup will be pursued to cover the 13 remaining Tier-2 clusters that lack ground-based follow-up.

%%%%%%%%%%%%%%%%

\subsection{\cxo\ X-ray}

Accompanying \cxo\ data for the CHEX-MATE clusters will be of importance in the completion of certain project goals. In particular, its high spatial resolution is preferred for studying the central regions of clusters (within ~100 kpc of the centre). This will be crucial when it comes to detecting the presence of cavities and other key AGN feedback features, along with studying and mapping the thermodynamic properties of the core. \cxo\ observations will also be used to detect and characterise point sources that are unresolved in the \xmm\ data (their expected variability in X-ray flux between observation epochs notwithstanding; \citealt{mau19}).

At the time of writing, 101/118 galaxy clusters in the sample have available \cxo\ data. Additionally, public data for PSZ2 G004.45-19.55 should be available soon, and PSZ2 G111.75+70.37 is within the field of view of a scheduled observation. However, the only available data for PSZ2 G067.52+34.75 (ObsID 14988) is unsuitable for galaxy cluster science, as not only is the observation limited to a single ACIS-S chip, but it also has a restrictive custom sub-array applied.

The \cxo\ coverage is representative of the full sample in mass and should be sufficient for the goals described above. In general, the data quality across the sample is good, with a minimum depth of >1600 counts (between 0.6 and 9.0 keV) within $\Rv$. This is comparable to the data quality used for cavity searches in \citet{hla15}. In the central 100 kpc, this translates to a median data quality of $\sim$1700 counts in the 0.7-2.0 keV energy band.

%%%%%%%%%%%%%%%%

\subsection{Radio}

Radio observations of galaxy clusters show several types of sources connected to the ICM (see \citealt{vanWeeren19} for a recent review). Radio halos are Mpc-size sources located at the cluster centres and are possibly due to turbulent re-acceleration during major mergers. Radio relics are arc-like radio sources located at the cluster periphery and linked to shock (re)accelerations. Mini-halos are sources of a few hundred kpc in size found at the centre of cool-core clusters surrounding the bright radio-loud BCG \citep{git18}. 

To understand the origin of radio halos and relics, it is important to quantify their occurrence as a function of cluster mass, redshift, and dynamical state. The CHEX-MATE samples represent good starting point for this analysis, which will complement the mass-complete samples already studied or planned \citep{cas13,cuc15}. Despite the number of archival observations in the radio band, the different sensitivity and observing bands of the clusters do not permit us to derive firm conclusions on the occurrence and evolution of radio halos and relics.
The fraction of clusters known to host a radio halo, relic, or mini halo in Tiers-1 and 2 are listed in Table \ref{tab:radio}. 
In the coming years, radio surveys with new and up-coming facilities will provide data with homogeneous sensitivity to cluster diffuse emission,  allowing one to perform unbiased statistical studies on the occurrence of radio halos, relics, and mini halos, and on their evolution with time.

%------------
% Radio table
%
\begin{table}
\centering
\begin{tabular}{lccc}
\toprule
\toprule
 Sample & Radio halos & Radio relics & Mini halos  \\
 \midrule
Tier-1  &   7\%        &    3\%       & 10\% \\           
Tier-2  &  39 \%       &   5\%        & 5\% \\
\bottomrule
\end{tabular}
\caption{\footnotesize Fraction of clusters with known diffuse radio sources for each Tier.}
\label{tab:radio}
\end{table}
%------------

Specifically, the Low Frequency Array (LOFAR) Two-metre Sky Survey (LoTSS, \citealt{Shimwell19}) will observe the Northern sky with unprecedented sensitivity ($\leq 100~\mu$Jy/beam) and resolution ($6"$) at low radio frequency 120-168 MHz, providing a complete view of non-thermal phenomena in galaxy clusters. 
All CHEX-MATE clusters at DEC$>0$; that is, 82 of 118 objects, would have a guaranteed LOFAR follow-up in the framework of LoTSS. Sixty clusters have already been observed by LoTSS at the time of writing.
In the Southern sky, other surveys are providing a homogeneous coverage of clusters. These include the GaLactic and Extragalactic All-sky MWA survey \citep[GLEAM, ][]{George17}, undertaken with the Murchison Widefield Array, and the Evolutionary Map of the Universe survey  \citep[EMU, ][]{Norris11}, undertaken with the Australian Square Kilometre Array Pathfinder. These 
will complement LoTSS with a similar resolution and sensitivity to extended cluster radio emission. The GLEAM survey (and EMU in the coming years) covers the entire sky south of DEC$>+30$ and is thus expected to provide a radio coverage of about 86 clusters. 

%%%%%%%%%%%%%%%%

\subsection{Hydrodynamical cluster simulations}\label{sec:sims}

In addition to the multi-wavelength observational data, theoretical input to CHEX-MATE will also be furnished with a large suite of hydrodynamical simulations of galaxy clusters, providing unprecedented statistics of these massive objects. The simulations are crucial for two main reasons. Firstly, they can be used for
interpreting the observational data to further our understanding of cluster physics; for example,  models of chemical enrichment,
stellar and black hole feedback, magnetic fields, and hydrodynamical processes such as viscosity, turbulence and conduction.
This will be achieved through comparison of observed and simulated cluster properties such as radial profiles (e.g. entropy,
temperature, pressure and metallicity) and global scaling relations between observables (e.g. X-ray luminosity, temperature,
SZE flux) and cluster mass within different apertures. For the latter, this will include mass estimates from simulated X-ray, SZE and lensing profiles, as well as their true values. Secondly, they are being used to
study the effects of cluster selection; for example, comparing clusters selected with SZE versus X-ray flux and assessing the impact of large-scale structure along the  line-of-sight, as well as allowing
simulated cluster samples with similar characteristics to the observed sample (e.g. in mass, redshift and morphology)
to be identified. We are also looking at related issues, such as  cluster centring, 
classifying clusters using various
dynamical and structural estimators, and investigating the level of hydrostatic mass bias (including how it is estimated, and how it depends on mass, redshift and dynamical state). 
%%%%%%%%%%%%%%%%

Simulation data are initially being provided using a number of existing data sets. In particular, we are using The Three Hundred \citep{cui18}, 
BAHAMAS+MACSIS \citep{mcc17,bar17a}
and Magneticum \citep{dol16}
simulations as these contain significant numbers of clusters that occupy the relevant
regions of mass-redshift space for both Tier-1 and Tier-2 samples (e.g. the largest Magneticum box contains over 200
thousand clusters in the Tier-1 mass range at redshift, $z=0$, and over 300 in the Tier-2 mass range at $z \simeq 0.5$).
These simulations are supplemented with a wide range of other runs available within the collaboration, which are also very
useful for addressing specific science projects using the CHEX-MATE data (e.g. \citealt{bar17b,bar18a,gas18,lebrun18,ras15,rup19,vaz17}). 
Beyond this, we will investigate the creation of  
bespoke simulated cluster samples for CHEX-MATE, taking into account both 
the latest cluster physics models and simulation codes available to the collaboration. High-resolution simulations will be also useful to generate detailed synthetic maps with different systematic and statistical errors and instrument responses.%%%%%%%%%%%%%%%%

%%%%%%%%%%%%%%%%
\section{Summary and conclusions}
\label{sec:conc}

The CHEX-MATE sample of 118 systems has been built as a future reference for clusters in the local volume and in the high mass regime. Its unique construction ensures that it contains not only the objects that make up the bulk of the population, but also the most massive systems, which are the most interesting targets for detailed multi-wavelength follow-up. The  project is intended to yield fundamental insights into the cluster mass scale and its relationship to the baryonic observables. It is conceived to be the key reference for numerical simulations, providing an observational calibration of the scaling laws between baryonic quantities and the underlying mass; it will provide the ultimate overview of the structural properties; and it will uncover the links between global and structural properties and the dynamical state and the presence of central cooling gas. 

A high-quality, homogeneous data set is critical in order to fulfil these objectives. We have detailed the X-ray observation preparation, exposure time calculation, and data analysis procedures needed to obtain the desired result, and we have shown that the new observations obtained for the project are in line with expectations. 
Although the X-ray observations are the backbone of the project, it is intrinsically multi-wavelength in nature. The majority of the sample is already covered by an  extremely rich data set comprising multi-band optical, SZE, and radio observations. Through its various working groups, the CHEX-MATE collaboration has embarked upon a considerable effort to completing this multi-wavelength follow-up. A parallel numerical simulation effort is also being undertaken.

The project legacy will be considerable. The sample corresponds to the descendants of the high–$z$ clusters that will be detected by upcoming SZE surveys such as SPT-3G, and the project will also provide key input for the interpretation of eROSITA survey data. Ultimately, we would like a method to detect clusters based on their most fundamental property: the total mass. This is becoming possible through WL analysis of the increasingly available high-quality, large-area, multi-band optical imaging data sets. Our project has particular synergy with {\it Euclid}, the sensitivity of which should allow blind detection of objects uniquely through their WL signal in the redshift and mass range covered by our sample. In the longer term, our sample will provide the targets of reference for dedicated Athena pointings for the deep exploration of ICM physics.

CHEX-MATE represents a very large investment of \xmm\ exposure time. The data are intended to be a community resource, and as such the X-ray observations do not have a proprietary period. They may be downloaded from the \xmm\ archive immediately after they have been obtained and processed by the \xmm\ SOC. This paper includes the first public release of the CHEX-MATE source list and X-ray observation details. Our hope is that the sample will be the foundation for cluster science with next-generation instruments for many years to come, fully justifying the investment in \xmm\ observing time and providing a unique heritage for ESA’s most successful astronomy mission.

%%%%%%%%%%%%%%%%

\begin{acknowledgements}

The results reported in this article are based on data obtained with \xmm, an ESA science mission with instruments and contributions directly funded by ESA Member States and NASA. 
We thank L. Ballo and XMM Science operation centre  for their extensive help in optimising  the observations. We thank N. Schartel and B. Wilkes for their support, particularly with  regard to  the joint \cxo-\xmm\ programme. 
\planck\ (\url{www.esa.int/Planck}) was an ESA project with instruments provided by two scientific consortia funded by ESA member states (in particular the lead countries France and Italy), with contributions from NASA (USA) and telescope reflectors provided by a collaboration between ESA and a scientific consortium led and funded by Denmark.
The scientific results reported in this article are based in part on observations made by the \cxo\ X-ray Observatory. 
This research has made use of the Science Analysis Software (SAS) provided by the XMM SOC and  the Chandra X-ray Center (CXC)  application packages CIAO, ChIPS, and Sherpa.
M.A., G.W.P, I.B, H.A., J-B.M., A.M.C.L., P.T.. S.Z. acknowledge funding from the European Research  Council  under  the  European  Union’s  Seventh  Framework Programme (FP72007-2013) ERC grant agreement no 340519. This work has benefitted from CNES and PNCG funding and A.I. acknowledges support from the CNES fellowship program. 
S.E., M.S., H.B., F.G., M.R., L.L., S.B., I.B., S.D.G., S.G., P.M., S.M., E.R. acknowledge financial contribution from the contracts 
% ASI-INAF Athena 2015-046-R.0, 
ASI-INAF Athena 2019-27-HH.0,
`Attivit\`a di Studio per la comunit\`a scientifica di Astrofisica delle Alte Energie e Fisica Astroparticellare' (Accordo Attuativo ASI-INAF n. 2017-14-H.0), and from INAF mainstream project 1.05.01.86.10.
S.E. acknowledges support from the European Union’s Horizon 2020 Programme under the AHEAD2020 project (grant agreement n. 871158).
B.J.M. and R.T.D. acknowledge support from STFC grant ST/R000700/1.
M.N. is partly supported by INAF-1.05.01.86.20.
A.S., L.S. and B.S. are supported by the ERC-StG `ClustersXCosmo' grant agreement 716762, and by the FARE-MIUR grant `ClustersXEuclid' R165SBKTMA.
A.B. acknowledges support from the ERC Starting Grant `DRANOEL' n.714245 and from the MIUR grant FARE `SMS'.
S.B. acknowledges also support from the INFN InDark Grant.
K.D. acknowledges support by the Deutsche Forschungsgemeinschaft (DFG, German Research Foundation) under Germany's Excellence Strategy -- EXC-2094 -- 3907833.
M.D. acknowledges NASA ADAP/SAO Award SV9-89010.
M.D.P. acknowledges support from Sapienza Universit\'a di Roma thanks to Progetti di Ricerca Medi 2019, prot. RM11916B7540DD8D.
M.G. acknowledges support from NASA Chandra GO8-19104X/GO9-20114X and HST GO-15890.020-A.
M.J. is supported by the United Kingdom Research and Innovation (UKRI) Future Leaders Fellowship `Using Cosmic Beasts to uncover the Nature of Dark Matter' (grant number MR/S017216/1).
F.K. acknowledges support by the French National Research Agency in the framework of the `Investissements d’avenir  program (ANR-15-IDEX-02)'.
A.M.C.L.B. acknowledges also funding from a PSL  University Research Fellowship.
J.A.R.-M. acknowledges support by the Spanish Ministry of Science  and Innovation (MICINN) under the projects AYA2014-60438-P and AYA2017-84185-P.
K.U. acknowledges support from the Ministry of Science and Technology of Taiwan (grant MOST 109-2112-M-001-018-MY3) and Academia Sinica (grant ASIA-107-M01).
F.V. acknowledges financial support from the ERC Starting Grant `MAGCOW', no.714196.
G.Y. acknowledges financial support by  MICIU/FEDER (Spain) under project grant PGC2018-094975-C21.
CHEX-MATE has benefitted from support from the International Space Science Institute (ISSI) in Bern and we acknowledge ISSI's hospitality.
This research made use of a number of \texttt{python} packages, including: \texttt{astropy} \citep{astropy:2018}, \texttt{matplotlib} \citep{Hunter2007}, \texttt{numpy} \citep{Numpy2011}, and  \texttt{scipy} \citep{Jones2001}.
\end{acknowledgements}

\bibliographystyle{aa} 
\bibliography{heritage_pres} 

\appendix

%%%%%%%%%%%%%%

\section{Sample selection strategy}
\label{sec:appx}

\subsection{Detailed selection strategy}

\begin{figure*}[!h]
\centering
\includegraphics[bb=50 115 540 810 ,clip,scale=1.,angle=0,keepaspectratio,width=0.90\textwidth]{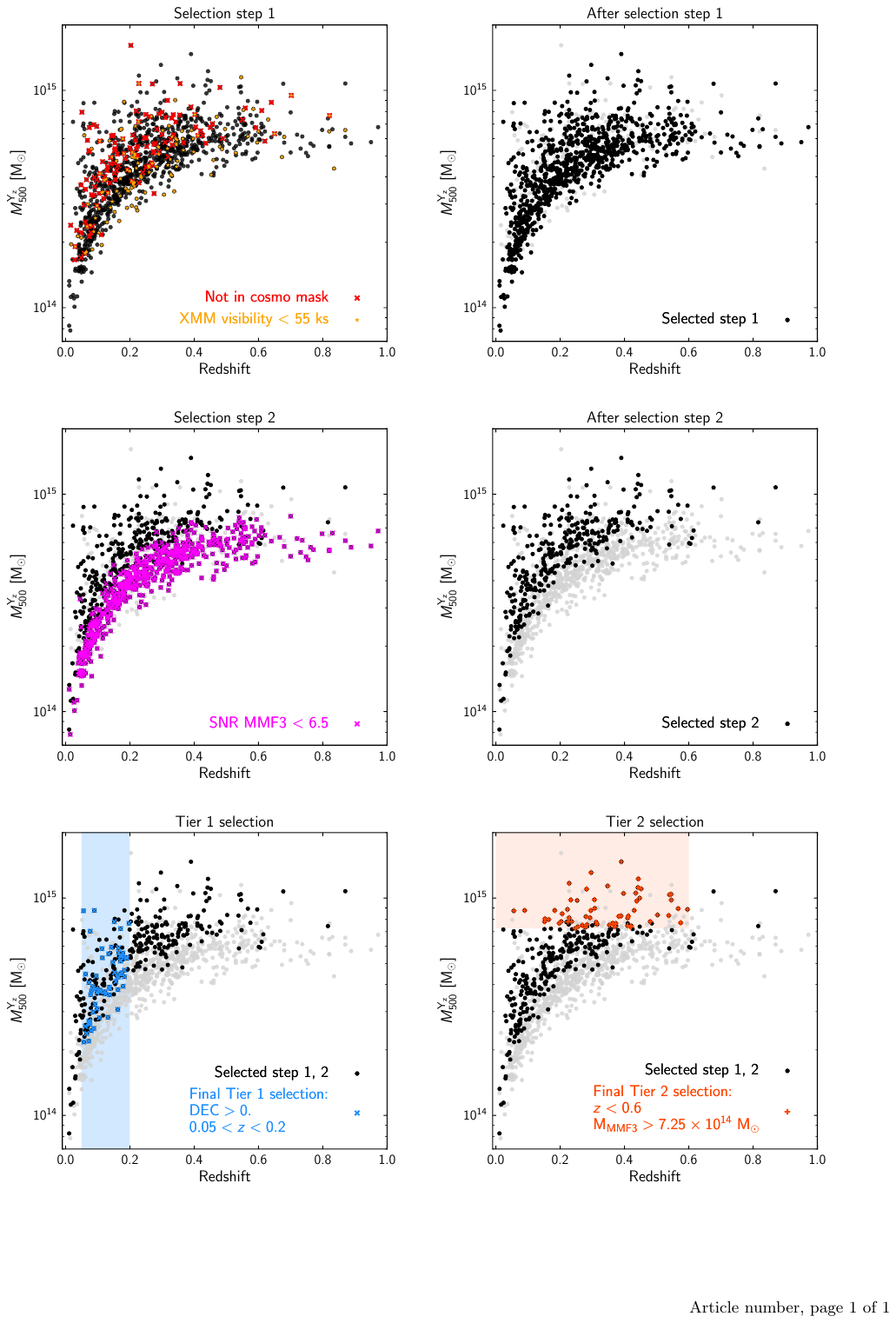}
\caption{\footnotesize Sample selection strategy. See Appendix~\ref{sec:appx} for a detailed explanation.}
\label{fig:appx}
\end{figure*}

Figure~\ref{fig:appx} illustrates the sample selection strategy. Starting from the PSZ2 catalogue, clusters that were successively removed owing to various criteria are marked in the $z$--$M_{500}$ plane, where $M_{500}$ is the PSZ2 mass. 

The first selection step is based on criteria that depend on source position on the sky, and is illustrated in the top panels. Sources outside the \planck\ cosmological mask, or with \xmm\ visibility less than 55 ksec were first excluded. Those with known redshift are marked by red crosses and orange points, respectively, in the top-left panel. The clusters remaining after this selection are shown with black points in the right panel; excluded clusters are shown in grey.  We note that the \xmm\ visibility criterion depends only on the object position in the sky. The region excluded by the visibility criterion is shown as a pink shaded area in Fig.~\ref{fig:lens}.

The second selection step is based on a criterion that depends on the S/N of the \planck\ SZE detection by the MMF3 algorithm. In the left-middle panel, sources that met the previous criteria but that are at ${\rm S/N_{MMF3}} < 6.5$ are marked with magenta crosses. Objects excluded in this and the first step are shown in grey in the middle-right panel. The remaining clusters are marked with black points; these, plus two not yet validated cluster candidates, comprise the parent cluster sample used for the definition of Tier-1 and Tier-2.
 
The bottom-left panel identifies the Tier-1 clusters with blue crosses. These are the clusters of the parent sample (black points) that are in the redshift range $0.05 < z < 0.2$, delimited by the blue shaded region, and in the northern sky (${\rm Dec} > 0$). Objects in the blue region but not marked in blue correspond to clusters in the southern sky. 

The bottom-right panel identifies the Tier-2 clusters with red plusses. These are the clusters of the parent sample (black points) that are in the redshift range $ z < 0.6$ and with mass $M_{\rm 500, MMF3}$, derived from the MMF3 detection, greater than $7.25 \times 10^{14}\msol$. The red shaded area delimits the corresponding region in the  $z$--$M_{500}$ plane. The PSZ2 catalogue mass, $M_{500}$, was derived from the detection method that yielded the highest S/N. The nine clusters not marked by red plusses in the orange area have been detected at higher S/N by  methods other than MMF3, but have $M_{\rm 500, MMF3} < 7.25\times 10^{14}\msol$. 

%------------
% Appendix figure showing selection steps
%
%------------

%%%%%%%%%%%%%%

\subsection{Original selection}

The total exposure time available in the  original call for \xmm\ Multi-Year Heritage programmes was 6 Msec. Our proposal was one of two programmes selected, and the total time was divided equally between them.
Originally, we applied for 6 Msec to observe 169 (instead of 118) clusters. These objects were selected from the PSZ2 catalogue at S/N$>$6, and divided into two Tiers. Tier-1 consisted of 85 objects at $z<0.2$ with median $M_{500}$ of $3.7 \times 10^{14}\msol$. Tier-2 contained  84 systems at $z<0.6$ with $M_{500} > 7 \times 10^{14}\msol$. There were seven objects in common between the two Tiers. To accommodate the reduction in allocated exposure time, and maintain our ability to work with our original goal of a `large, unbiased, signal-to-noise-limited sample' we made several changes to the original object selection strategy. We raised the S/N cut to 6.5; we imposed a lower redshift cut of $z>0.05$ on the Tier-1 systems, which would have required deep off-axis observations to cover their emission up to $R_{500}$); we imposed a higher mass cut of $M_{500} > 7.25 \times 10^{14}\msol$ on the Tier-2 sample. 

After making these changes, the net reduction of 50\% in exposure time corresponded to a reduction of about 30\% in the number of objects.

%%%%%%%%%%%%%%%

\section{List of targets}
\label{appx:targetlist}

Tables~\ref{tab:master1b}, \ref{tab:master2b}, and \ref{tab:master3b} list the CHEX-MATE observations. They detail the following:
    the PSZ2 name; the coordinates of the X-ray peak in RA and Dec; the redshift; the nominal $M_{500}$ from the PSZ2 catalogue in units of $10^{14}\, \msol$; the signal-to-noise ratio (S/N); the Tier to which the object belongs (either 1 or 2, or `12' when the object belongs to both Tiers); the nominal Galactic absorption in units of $10^{20}\, {\rm cm}^{-2}$; the archived \xmm\ exposure time in kiloseconds; the archived \cxo\ exposure time in kiloseconds; the requested new \xmm\ exposure time in kiloseconds; the OBSid that identifies the observations used for the analysis (new exposures highlighted in bold font).

%------------
% List of targets table
%
\begin{sidewaystable*}
\centering
\begin{tabular}{c@{\hskip5.pt} c@{\hskip5.pt} c@{\hskip5.pt} c@{\hskip5.pt} c@{\hskip2.pt} c@{\hskip2.pt} c@{\hskip2.pt} c@{\hskip2.pt} c@{\hskip2.pt} c@{\hskip2.pt} c@{\hskip2.pt} l} 
\toprule
\toprule
     Name & RA    & Dec  & $z$ & $M_{500}$ & SNR & Tier & $n_H$ & $t_{\rm XMM}$ & $t_{\rm CXO}$ & $t_{\rm XMM, new}$ & OBSid \\
      & h:m:s & d:m:s &    & $10^{14} M_{\odot}$ & & & $10^{20}$ cm$^{-2}$ & ksec & ksec & ksec & \\ 
       \midrule

PSZ2 G000.13+78.04 & 13:34:08.2 & +20:14:26 & 0.1710 & 5.1 & 9.3 & 1 & 1.8 & 0.0 & 7.1 & 23.0 & {\scriptsize   0821810801} \\
PSZ2 G004.45-19.55 & 19:17:04.6 & -33:31:20 & 0.5400 & 10.1 & 9.1 & 2 & 5.9 & 10.0 & 0.0 & 48.0 & {\scriptsize   0656201001} \\
PSZ2 G006.49+50.56 & 15:10:56.2 & +05:44:42 & 0.0766 & 7.0 & 23.2 & 1 & 3.2 & 176.5 & 128.5 & 0.0 & {\scriptsize   0111270201 055178(02-05)01 074441(09-12)01} \\
PSZ2 G008.31-64.74 & 22:58:48.3 & -34:47:59 & 0.3120 & 7.4 & 10.9 & 2 & 1.3 & 0.0 & 73.5 & 31.0 & {\scriptsize   {\bf 0827010901}} \\
PSZ2 G008.94-81.22 & 00:14:19.0 & -30:23:09 & 0.3066 & 9.0 & 15.3 & 2 & 1.4 & 119.7 & 126.8 & 0.0 & {\scriptsize   0042340101 0743850101} \\
PSZ2 G021.10+33.24 & 16:32:48.0 & +05:34:30 & 0.1514 & 7.8 & 16.3 & 12 & 5.7 & 105.6 & 118.3 & 0.0 & {\scriptsize   0112230301 030649(01-04)01} \\
PSZ2 G028.63+50.15 & 15:40:09.1 & +17:52:40 & 0.0916 & 3.2 & 6.9 & 1 & 2.6 & 0.0 & 0.0 & 21.0 & {\scriptsize   0821810401} \\
PSZ2 G028.89+60.13 & 15:00:19.7 & +21:22:10 & 0.1530 & 4.5 & 7.5 & 1 & 3.4 & 21.7 & 20.1 & 21.0 & {\scriptsize   0693011001} \\
PSZ2 G031.93+78.71 & 13:41:48.8 & +26:22:22 & 0.0724 & 2.7 & 7.5 & 1 & 1.0 & 31.0 & 100.1 & 0.0 & {\scriptsize   0108460101} \\
PSZ2 G033.81+77.18 & 13:48:52.9 & +26:35:30 & 0.0622 & 4.5 & 19.3 & 1 & 1.2 & 117.9 & 289.9 & 0.0 & {\scriptsize   0097820101 020519(01-02)01 0744412001 0744412101} \\
PSZ2 G040.03+74.95 & 13:59:14.8 & +27:58:33 & 0.0612 & 2.3 & 9.3 & 1 & 1.3 & 0.0 & 10.1 & 21.0 & {\scriptsize   {\bf 0827030901}} \\
PSZ2 G040.03+74.95 N & 13:59:09.0 & +28:06:48 & 0.0612 & 2.3 & 9.3 & 1 & 1.3 & 0.0 & 10.1 & 21.0 & {\scriptsize   {\bf 0827030801}} \\
PSZ2 G040.58+77.12 & 13:49:23.8 & +28:06:32 & 0.0748 & 2.6 & 7.6 & 1 & 1.1 & 0.0 & 11.1 & 21.0 & {\scriptsize   {\bf 0827041901} {\bf 0827340601}} \\
PSZ2 G041.45+29.10 & 17:17:45.4 & +19:40:38 & 0.1780 & 5.4 & 9.4 & 1 & 4.7 & 57.1 & 0.0 & 0.0 & {\scriptsize   0601080101} \\
PSZ2 G042.81+56.61 & 15:22:29.3 & +27:42:22 & 0.0723 & 4.2 & 12.5 & 1 & 3.0 & 49.9 & 55.2 & 0.0 & {\scriptsize   0202080201} \\
PSZ2 G042.81+56.61 NW & 15:21:53.2 & +27:52:56 & 0.0723 & 4.2 & 12.5 & 1 & 3.0 & 49.9 & 55.2 & 21.0 & {\scriptsize  } \\
PSZ2 G044.20+48.66 & 15:58:20.6 & +27:13:44 & 0.0894 & 8.8 & 28.4 & 12 & 3.8 & 83.2 & 404.1 & 0.0 & {\scriptsize   0111870301 0674560201 069444(01-02, 05-06)01} \\
PSZ2 G044.77-51.30 & 22:14:57.5 & -14:00:14 & 0.5027 & 8.4 & 8.3 & 2 & 2.9 & 19.5 & 38.5 & 53.0 & {\scriptsize   0693661901} \\
PSZ2 G046.10+27.18 & 17:31:38.6 & +22:51:57 & 0.3890 & 7.8 & 9.0 & 2 & 5.0 & 16.1 & 20.8 & 37.0 & {\scriptsize   0723160601 {\bf 0827060201}} \\
PSZ2 G046.88+56.48 & 15:24:07.5 & +29:53:16 & 0.1145 & 5.1 & 12.9 & 1 & 1.9 & 0.0 & 56.2 & 21.0 & {\scriptsize   {\bf 0827010601}} \\
PSZ2 G048.10+57.16 & 15:21:13.8 & +30:38:32 & 0.0777 & 3.5 & 9.9 & 1 & 1.7 & 44.7 & 31.8 & 0.0 & {\scriptsize   0721740101} \\
PSZ2 G049.22+30.87 & 17:20:09.8 & +26:37:31 & 0.1644 & 5.9 & 11.8 & 1 & 3.4 & 68.0 & 60.5 & 0.0 & {\scriptsize   050067(02-04)01} \\
PSZ2 G049.32+44.37 & 16:20:30.4 & +29:53:36 & 0.0972 & 3.8 & 8.0 & 1 & 2.6 & 10.0 & 12.1 & 21.0 & {\scriptsize   0692930901} \\
PSZ2 G050.40+31.17 & 17:20:08.2 & +27:40:07 & 0.1640 & 4.2 & 7.3 & 1 & 3.3 & 13.0 & 10.1 & 33.0 & {\scriptsize   {\bf 0827040101}} \\
PSZ2 G053.53+59.52 & 15:10:12.5 & +33:30:37 & 0.1130 & 5.2 & 15.9 & 1 & 1.6 & 63.5 & 290.9 & 0.0 & {\scriptsize   0149880101 0303930101} \\
PSZ2 G055.59+31.85 & 17:22:27.4 & +32:07:55 & 0.2240 & 7.7 & 14.0 & 2 & 3.2 & 60.6 & 133.8 & 0.0 & {\scriptsize   0093030301 0093031001 0693180901} \\
PSZ2 G056.77+36.32 & 17:02:42.6 & +34:03:36 & 0.0953 & 4.3 & 12.7 & 1 & 1.9 & 22.7 & 62.9 & 0.0 & {\scriptsize   0740900101} \\
PSZ2 G056.93-55.08 & 22:43:21.4 & -09:35:42 & 0.4470 & 9.5 & 11.8 & 2 & 4.0 & 128.6 & 120.9 & 0.0 & {\scriptsize   0503490201} \\
PSZ2 G057.25-45.34 & 22:11:45.8 & -03:49:47 & 0.3970 & 9.6 & 12.5 & 2 & 5.5 & 29.7 & 18.0 & 21.0 & {\scriptsize   0693010601} \\
PSZ2 G057.61+34.93 & 17:09:49.2 & +34:27:11 & 0.0802 & 3.7 & 11.6 & 1 & 2.2 & 0.0 & 10.1 & 21.0 & {\scriptsize   {\bf 0827010501}} \\
PSZ2 G057.78+52.32 & 15:44:58.9 & +36:06:30 & 0.0654 & 2.3 & 7.2 & 1 & 1.7 & 0.0 & 19.6 & 21.0 & {\scriptsize   {\bf 0827040301}} \\
PSZ2 G057.78+52.32 N & 15:44:54.1 & +36:19:34 & 0.0654 & 2.3 & 7.2 & 1 & 1.7 & 0.0 & 19.6 & 21.0 & {\scriptsize   {\bf 0827040201}$^{\star}$} \\
PSZ2 G057.92+27.64 & 17:44:14.5 & +32:59:28 & 0.0757 & 2.7 & 8.0 & 1 & 3.8 & 0.0 & 54.2 & 21.0 & {\scriptsize   {\bf 0827030301}} \\
PSZ2 G062.46-21.35 & 21:04:53.2 & +14:01:27 & 0.1615 & 4.1 & 7.0 & 1 & 7.1 & 0.0 & 0.0 & 40.0 & {\scriptsize   {\bf 0827040701}} \\
PSZ2 G066.41+27.03 & 17:56:50.4 & +40:08:02 & 0.5750 & 7.7 & 8.8 & 2 & 3.4 & 0.0 & 12.8 & 105.0 & {\scriptsize   {\bf 0827320601}} \\
PSZ2 G066.68+68.44 & 14:21:40.6 & +37:17:29 & 0.1630 & 3.8 & 7.1 & 1 & 1.0 & 0.0 & 5.1 & 38.0 & {\scriptsize   {\bf 0827031401}} \\
PSZ2 G067.17+67.46 & 14:26:02.3 & +37:49:27 & 0.1712 & 7.1 & 17.8 & 1 & 1.1 & 22.2 & 169.7 & 0.0 & {\scriptsize   0112230201} \\
PSZ2 G067.52+34.75 & 17:17:19.2 & +42:26:58 & 0.1754 & 4.5 & 8.4 & 1 & 1.6 & 0.0 & 20.1 & 31.0 & {\scriptsize   {\bf 0827021101}$^{\star}$} \\
PSZ2 G068.22+15.18 & 18:57:37.7 & +38:00:32 & 0.0567 & 2.1 & 8.2 & 1 & 7.2 & 7.7 & 9.7 & 21.0 & {\scriptsize   0762800801 {\bf 0827041201}} \\
PSZ2 G068.22+15.18 E & 18:57:33.4 & +38:01:09 & 0.0567 & 2.1 & 8.2 & 1 & 7.2 & 7.7 & 9.7 & 21.0 & {\scriptsize   {\bf 0827041101}} \\
\bottomrule
\end{tabular}
    \caption{\footnotesize List of 
    CHEX-MATE \xmm\ observations. We quote:
    the PSZ2 name; the coordinates of the X-ray peak; the redshift; the nominal $M_{500}$ from the PSZ2 catalogue; the Signal-to-Noise ratio; the Tier to which the object belongs (either 1 or 2; "12" when the object is part of both Tiers); the nominal Galactic absorption; the archived \xmm\ exposure time; the archived \cxo\ exposure time; the requested new \xmm\ exposure time; the OBSid that identifies the observations used for the analysis (in bold font, the new exposures available on September 9 2020; the symbol $^{\star}$ identifies the targets that will be re-observed in the final year). }
    \label{tab:master1b}
\end{sidewaystable*}
%------------

\begin{sidewaystable*}
    \centering
     \begin{tabular}{c@{\hskip5.pt} c@{\hskip5.pt} c@{\hskip5.pt} c@{\hskip5.pt} c@{\hskip2.pt} c@{\hskip2.pt} c@{\hskip2.pt} c@{\hskip2.pt} c@{\hskip2.pt} c@{\hskip2.pt} c@{\hskip2.pt} l} 
    %  \hline \hline
\toprule
\toprule
     Name & RA    & Dec  & $z$ & $M_{500}$ & SNR & Tier & $n_H$ & $t_{\rm XMM}$ & $t_{\rm CXO}$ & $t_{\rm XMM, new}$ & OBSid \\
      & h:m:s & d:m:s &    & $10^{14} M_{\odot}$ & & & $10^{20}$ cm$^{-2}$ & ksec & ksec & ksec & \\ 
      \midrule
PSZ2 G071.63+29.78 & 17:47:14.0 & +45:11:45 & 0.1565 & 4.1 & 8.7 & 1 & 2.7 & 19.0 & 22.6 & 22.0 & {\scriptsize   0692932201 {\bf 0827050501}} \\
PSZ2 G072.62+41.46 & 16:40:19.9 & +46:42:39 & 0.2280 & 11.4 & 27.4 & 2 & 1.8 & 40.1 & 497.4 & 0.0 & {\scriptsize   0112231801 0112231901 0605000501} \\
PSZ2 G073.97-27.82 & 21:53:36.8 & +17:41:41 & 0.2329 & 9.5 & 19.1 & 2 & 6.2 & 20.0 & 115.4 & 21.0 & {\scriptsize   0111270101} \\
PSZ2 G075.71+13.51 & 19:21:12.0 & +43:56:49 & 0.0557 & 8.7 & 49.0 & 12 & 8.1 & 81.9 & 90.8 & 0.0 & {\scriptsize   030215(01-02)01 0600040101 0743840201 074441(01-03)01 0763490301} \\
PSZ2 G077.90-26.63 & 22:00:52.5 & +20:58:04 & 0.1470 & 5.0 & 11.1 & 1 & 6.6 & 0.0 & 10.4 & 23.0 & {\scriptsize   {\bf 0827020101}} \\
PSZ2 G080.16+57.65 & 15:01:07.9 & +47:16:35 & 0.0878 & 2.5 & 7.8 & 1 & 2.5 & 0.0 & 0.0 & 29.0 & {\scriptsize   0821040401$^{\star}$} \\
PSZ2 G080.37+14.64 & 19:26:10.0 & +48:33:10 & 0.0980 & 3.1 & 8.1 & 1 & 6.9 & 0.0 & 0.0 & 29.0 & {\scriptsize   {\bf 0827031501}$^{\star}$} \\
PSZ2 G080.41-33.24 & 22:26:06.1 & +17:21:45 & 0.1072 & 3.8 & 8.7 & 1 & 5.1 & 108.8 & 124.8 & 0.0 & {\scriptsize   0762470101} \\
PSZ2 G083.29-31.03 & 22:28:33.7 & +20:37:16 & 0.4120 & 7.6 & 9.5 & 2 & 4.3 & 22.3 & 20.1 & 30.0 & {\scriptsize   0147890101 {\bf 0827360901}} \\
PSZ2 G083.86+85.09 & 13:05:51.2 & +30:53:42 & 0.1832 & 4.7 & 8.0 & 1 & 1.0 & 0.0 & 0.0 & 31.0 & {\scriptsize   {\bf 0827030701}} \\
PSZ2 G085.98+26.69 & 18:19:54.2 & +57:09:22 & 0.1790 & 4.2 & 9.3 & 1 & 3.9 & 35.0 & 34.8 & 42.0 & {\scriptsize   0692932701 {\bf 0827041601}} \\
PSZ2 G087.03-57.37 & 23:37:37.5 & +00:16:07 & 0.2779 & 7.3 & 11.2 & 2 & 3.5 & 9.0 & 26.4 & 22.0 & {\scriptsize   0042341301} \\
PSZ2 G092.71+73.46 & 13:35:18.2 & +41:00:06 & 0.2279 & 8.0 & 17.6 & 2 & 0.8 & 23.1 & 95.9 & 21.0 & {\scriptsize   0084230901} \\
PSZ2 G094.69+26.36 & 18:32:30.9 & +64:49:53 & 0.1623 & 3.1 & 7.7 & 1 & 4.4 & 19.7 & 0.0 & 53.0 & {\scriptsize   0762801101} \\
PSZ2 G098.44+56.59 & 14:27:25.4 & +55:44:27 & 0.1318 & 2.8 & 6.6 & 1 & 1.3 & 0.0 & 0.0 & 48.0 & {\scriptsize  } \\
PSZ2 G099.48+55.60 & 14:28:38.0 & +56:51:35 & 0.1051 & 2.7 & 8.1 & 1 & 1.3 & 0.0 & 15.0 & 30.0 & {\scriptsize   0821810901} \\
PSZ2 G105.55+77.21 & 13:11:08.7 & +39:13:37 & 0.0720 & 2.2 & 6.9 & 1 & 1.2 & 0.0 & 26.8 & 24.0 & {\scriptsize   {\bf 0827031301}} \\
PSZ2 G106.87-83.23 & 00:43:24.6 & -20:37:16 & 0.2924 & 7.7 & 12.7 & 2 & 1.8 & 10.0 & 405.6 & 21.0 & {\scriptsize   0042340201 {\bf 0827041501}} \\
PSZ2 G107.10+65.32 & 13:32:45.4 & +50:32:05 & 0.2799 & 7.8 & 16.5 & 2 & 1.0 & 55.3 & 235.6 & 0.0 & {\scriptsize   0142860201} \\
PSZ2 G111.61-45.71 & 00:18:33.2 & +16:26:10 & 0.5456 & 8.5 & 9.7 & 2 & 4.0 & 32.3 & 68.3 & 26.0 & {\scriptsize   0111000101 0111000201 {\bf 0827061301}} \\
PSZ2 G111.75+70.37 & 13:13:06.4 & +46:16:51 & 0.1830 & 4.3 & 8.3 & 1 & 1.1 & 0.0 & 0.0 & 36.0 & {\scriptsize   {\bf 0827020801}} \\
PSZ2 G113.29-29.69 & 00:11:45.9 & +32:24:51 & 0.1073 & 3.6 & 8.1 & 1 & 4.7 & 0.0 & 8.1 & 21.0 & {\scriptsize   {\bf 0827021201}} \\
PSZ2 G113.91-37.01 & 00:19:41.7 & +25:18:05 & 0.3712 & 7.6 & 8.2 & 2 & 2.9 & 0.0 & 0.0 & 49.0 & {\scriptsize   {\bf 0827021001}} \\
PSZ2 G114.79-33.71 & 00:20:37.5 & +28:39:32 & 0.0940 & 3.8 & 9.4 & 1 & 3.9 & 0.0 & 14.1 & 21.0 & {\scriptsize   {\bf 0827320401}} \\
PSZ2 G124.20-36.48 & 00:55:50.6 & +26:24:36 & 0.1971 & 7.3 & 12.8 & 12 & 5.4 & 40.7 & 366.3 & 0.0 & {\scriptsize   0203220101} \\
PSZ2 G143.26+65.24 & 11:59:14.8 & +49:47:33 & 0.3634 & 7.3 & 10.2 & 2 & 2.2 & 0.0 & 22.1 & 45.0 & {\scriptsize   {\bf 0827020201} {\bf 0827320201}} \\
PSZ2 G149.39-36.84 & 02:21:33.6 & +21:21:42 & 0.1700 & 5.3 & 7.3 & 1 & 7.5 & 0.0 & 0.0 & 28.0 & {\scriptsize   {\bf 0827030601}} \\
PSZ2 G155.27-68.42 & 01:37:24.8 & -08:27:20 & 0.5670 & 8.4 & 8.0 & 2 & 3.5 & 24.9 & 0.0 & 51.0 & {\scriptsize   06936628(01-02) 0700180201 {\bf 0827060801}} \\
PSZ2 G159.91-73.50 & 01:31:52.6 & -13:36:43 & 0.2060 & 8.5 & 17.1 & 2 & 1.4 & 63.7 & 20.2 & 0.0 & {\scriptsize   0084230301 0550960101} \\
PSZ2 G172.74+65.30 & 11:11:40.4 & +40:50:14 & 0.0794 & 2.4 & 7.5 & 1 & 1.3 & 0.0 & 11.7 & 23.0 & {\scriptsize   {\bf 0827031101}} \\
PSZ2 G172.98-53.55 & 02:39:53.3 & -01:34:44 & 0.3730 & 7.4 & 7.6 & 2 & 3.0 & 130.0 & 96.3 & 0.0 & {\scriptsize   0782150101} \\
PSZ2 G179.09+60.12 & 10:40:44.5 & +39:57:11 & 0.1372 & 3.8 & 7.5 & 1 & 1.7 & 36.6 & 58.1 & 27.0 & {\scriptsize   0147630101 {\bf 0827330401}} \\
PSZ2 G186.37+37.26 & 08:42:57.2 & +36:22:03 & 0.2820 & 11.0 & 18.9 & 2 & 2.9 & 22.4 & 27.9 & 21.0 & {\scriptsize   0605000701 {\bf 0827041001}} \\
PSZ2 G187.53+21.92 & 07:32:20.4 & +31:37:59 & 0.1710 & 5.2 & 8.1 & 1 & 4.9 & 18.6 & 217.6 & 21.0 & {\scriptsize   0673850201} \\
PSZ2 G192.18+56.12 & 10:16:21.9 & +33:38:26 & 0.1240 & 3.6 & 7.8 & 1 & 1.5 & 0.0 & 5.1 & 25.0 & {\scriptsize   0821810701} \\
PSZ2 G195.75-24.32 & 04:54:06.5 & +02:54:25 & 0.2030 & 7.8 & 11.8 & 2 & 5.7 & 42.4 & 535.2 & 0.0 & {\scriptsize   0201510101} \\
PSZ2 G201.50-27.31 & 04:54:11.1 & -03:00:58 & 0.5377 & 8.3 & 7.1 & 2 & 3.9 & 42.6 & 58.8 & 56.0 & {\scriptsize   0205670101 {\bf 0827061001}} \\
PSZ2 G204.10+16.51 & 07:35:47.5 & +15:06:48 & 0.1220 & 3.7 & 6.7 & 1 & 5.0 & 0.0 & 0.0 & 28.0 & {\scriptsize   {\bf 0827040401}} \\
PSZ2 G205.93-39.46 & 04:17:34.1 & -11:54:18 & 0.4430 & 11.5 & 13.8 & 2 & 3.3 & 0.0 & 90.3 & 26.0 & {\scriptsize   {\bf 0827010101} {\bf 0827011501} {\bf 0827310101}} \\
PSZ2 G206.45+13.89 & 07:29:50.7 & +11:56:28 & 0.4100 & 7.5 & 7.3 & 2 & 5.8 & 25.0 & 0.0 & 51.0 & {\scriptsize   0762440301} \\
\bottomrule
    % \hline
    \end{tabular}
    \caption{\footnotesize List of  \xmm\ observations.}
    \label{tab:master2b}
\end{sidewaystable*}
%------------
\begin{sidewaystable*}
    \centering
     \begin{tabular}{c@{\hskip5.pt} c@{\hskip5.pt} c@{\hskip5.pt} c@{\hskip5.pt} c@{\hskip2.pt} c@{\hskip2.pt} c@{\hskip2.pt} c@{\hskip2.pt} c@{\hskip2.pt} c@{\hskip2.pt} c@{\hskip2.pt} l} 
    %  \hline \hline
\toprule
\toprule
     Name & RA    & Dec  & $z$ & $M_{500}$ & SNR & Tier & $n_H$ & $t_{\rm XMM}$ & $t_{\rm CXO}$ & $t_{\rm XMM, new}$ & OBSid \\
      & h:m:s & d:m:s &    & $10^{14} M_{\odot}$ & & & $10^{20}$ cm$^{-2}$ & ksec & ksec & ksec & \\ \midrule
PSZ2 G207.88+81.31 & 12:12:18.2 & +27:33:05 & 0.3530 & 7.4 & 10.1 & 2 & 1.7 & 0.0 & 14.8 & 43.0 & {\scriptsize   {\bf 0827020301}} \\
PSZ2 G208.80-30.67 & 04:54:06.7 & -10:13:08 & 0.2475 & 7.3 & 9.5 & 2 & 4.8 & 85.5 & 168.6 & 0.0 & {\scriptsize   0603890101} \\
PSZ2 G210.64+17.09 & 07:48:46.6 & +09:40:00 & 0.4800 & 7.8 & 7.6 & 2 & 2.7 & 11.7 & 0.0 & 60.0 & {\scriptsize   0658200501} \\
PSZ2 G216.62+47.00 & 09:49:51.5 & +17:07:09 & 0.3826 & 8.5 & 9.8 & 2 & 3.1 & 12.7 & 35.6 & 43.0 & {\scriptsize   0723160201} \\
PSZ2 G217.09+40.15 & 09:24:05.9 & +14:10:26 & 0.1357 & 3.9 & 7.0 & 1 & 3.3 & 0.0 & 30.1 & 28.0 & {\scriptsize   {\bf 0827031001}} \\
PSZ2 G217.40+10.88 & 07:38:18.4 & +01:02:15 & 0.1890 & 5.3 & 7.8 & 1 & 8.5 & 0.0 & 0.0 & 37.0 & {\scriptsize   {\bf 0827330201}$^{\star}$} \\
PSZ2 G218.59+71.31 & 11:29:54.5 & +23:48:14 & 0.1371 & 3.8 & 7.2 & 1 & 1.2 & 0.0 & 0.0 & 30.0 & {\scriptsize  } \\
PSZ2 G218.81+35.51 & 09:09:12.6 & +10:58:32 & 0.1751 & 5.2 & 8.4 & 1 & 3.2 & 27.4 & 34.9 & 21.0 & {\scriptsize   0605000901 0673850901} \\
PSZ2 G224.00+69.33 & 11:23:57.6 & +21:28:56 & 0.1904 & 5.1 & 8.8 & 1 & 1.6 & 0.0 & 20.1 & 27.0 & {\scriptsize   {\bf 0827020901}} \\
PSZ2 G225.93-19.99 & 06:00:08.1 & -20:08:07 & 0.4350 & 9.8 & 12.4 & 2 & 4.9 & 9.0 & 0.0 & 26.0 & {\scriptsize   0650381401 {\bf 0827050601}} \\
PSZ2 G226.18+76.79 & 11:55:17.9 & +23:24:18 & 0.1427 & 6.0 & 14.6 & 1 & 1.8 & 279.8 & 157.9 & 0.0 & {\scriptsize   050269(01-02)01 055128(01-02)01} \\
PSZ2 G228.16+75.20 & 11:49:35.1 & +22:24:08 & 0.5450 & 9.8 & 11.4 & 2 & 1.9 & 24.7 & 370.2 & 42.0 & {\scriptsize   0693661701 {\bf 0827341301}} \\
PSZ2 G229.74+77.96 & 12:01:13.2 & +23:06:21 & 0.2690 & 7.4 & 12.0 & 2 & 2.2 & 0.0 & 26.1 & 27.0 & {\scriptsize   0821810501$^{\star}$} \\
PSZ2 G238.69+63.26 & 11:12:54.3 & +13:26:05 & 0.1690 & 4.2 & 7.3 & 1 & 1.6 & 50.0 & 93.2 & 0.0 & {\scriptsize   0500760101} \\
PSZ2 G239.27-26.01 & 05:53:28.4 & -33:42:33 & 0.4300 & 8.8 & 12.2 & 2 & 3.3 & 0.0 & 85.1 & 49.0 & {\scriptsize   {\bf 0827010401}$^{\star}$} \\
PSZ2 G241.11-28.68 & 05:42:57.1 & -35:59:49 & 0.4200 & 7.4 & 9.0 & 2 & 3.0 & 10.0 & 0.0 & 55.0 & {\scriptsize   0656202001 {\bf 0827050801}} \\
PSZ2 G243.15-73.84 & 01:59:02.7 & -34:12:57 & 0.4100 & 8.1 & 10.3 & 2 & 1.5 & 0.0 & 19.5 & 48.0 & {\scriptsize   {\bf 0827011301}} \\
PSZ2 G243.64+67.74 & 11:32:51.9 & +14:27:11 & 0.0834 & 3.6 & 12.0 & 1 & 2.9 & 0.0 & 9.1 & 21.0 & {\scriptsize   {\bf 0827010801}} \\
PSZ2 G259.98-63.43 & 02:32:18.7 & -44:20:46 & 0.2836 & 7.5 & 13.0 & 2 & 1.7 & 10.0 & 23.7 & 21.0 & {\scriptsize   0042340301} \\
PSZ2 G262.27-35.38 & 05:16:37.4 & -54:30:10 & 0.2952 & 8.8 & 22.9 & 2 & 2.1 & 58.1 & 31.4 & 0.0 & {\scriptsize   0042340701 0205330301 0692934301} \\
PSZ2 G262.73-40.92 & 04:38:17.4 & -54:19:23 & 0.4210 & 7.5 & 12.7 & 2 & 1.0 & 14.7 & 20.0 & 42.0 & {\scriptsize   0656201601} \\
PSZ2 G263.68-22.55 & 06:45:28.8 & -54:13:37 & 0.1644 & 8.0 & 21.7 & 2 & 5.6 & 47.4 & 10.1 & 0.0 & {\scriptsize   0201901201 0201903401 0404910401} \\
PSZ2 G266.04-21.25 & 06:58:30.0 & -55:56:23 & 0.2965 & 12.5 & 28.4 & 2 & 4.9 & 38.7 & 584.2 & 0.0 & {\scriptsize   0112980201} \\
PSZ2 G266.83+25.08 & 10:23:50.2 & -27:15:21 & 0.2542 & 7.3 & 11.7 & 2 & 5.6 & 0.0 & 37.2 & 29.0 & {\scriptsize   {\bf 0827011001}} \\
PSZ2 G271.18-30.95 & 05:49:19.6 & -62:05:13 & 0.3700 & 7.4 & 14.1 & 2 & 4.5 & 10.0 & 0.0 & 46.0 & {\scriptsize   0656201301 {\bf 0827050701}} \\
PSZ2 G273.59+63.27 & 12:00:25.4 & +03:20:49 & 0.1339 & 5.5 & 12.6 & 1 & 2.1 & 14.7 & 19.6 & 21.0 & {\scriptsize   {\bf 0827010301}} \\
PSZ2 G277.76-51.74 & 02:54:16.1 & -58:56:52 & 0.4380 & 8.7 & 15.1 & 2 & 1.9 & 68.6 & 0.0 & 0.0 & {\scriptsize   0656200301 0674380301} \\
PSZ2 G278.58+39.16 & 11:31:54.2 & -19:55:40 & 0.3075 & 8.3 & 12.2 & 2 & 4.0 & 10.0 & 99.2 & 21.0 & {\scriptsize   0042341001} \\
PSZ2 G283.91+73.87 & 10:16:21.9 & +33:38:26 & 0.0852 & 2.7 & 8.4 & 1 & 2.1 & 0.0 & 42.0 & 21.0 & {\scriptsize   {\bf 0827330501}} \\
PSZ2 G284.41+52.45 & 12:06:12.0 & -08:48:03 & 0.4414 & 10.4 & 13.6 & 2 & 4.3 & 178.9 & 203.8 & 0.0 & {\scriptsize   0502430401 0762070101} \\
PSZ2 G285.63+72.75 & 12:30:47.6 & +10:33:11 & 0.1650 & 5.6 & 10.9 & 1 & 2.1 & 0.0 & 19.2 & 21.0 & {\scriptsize   {\bf 0827011101}} \\
PSZ2 G286.98+32.90 & 11:50:49.0 & -28:04:28 & 0.3900 & 13.7 & 22.7 & 2 & 7.3 & 10.0 & 199.0 & 21.0 & {\scriptsize   0656201201 {\bf 0827341401}} \\
PSZ2 G287.46+81.12 & 12:41:17.6 & +18:34:28 & 0.0730 & 2.6 & 7.4 & 1 & 1.5 & 15.5 & 0.0 & 0.0 & {\scriptsize   0149900301} \\
PSZ2 G313.33+61.13 & 13:11:29.3 & -01:20:27 & 0.1832 & 8.8 & 16.7 & 2 & 1.8 & 84.8 & 199.8 & 0.0 & {\scriptsize   0093030101 069382(01-02)01} \\
PSZ2 G313.88-17.11 & 16:01:49.2 & -75:45:14 & 0.1530 & 7.9 & 16.6 & 2 & 5.9 & 31.5 & 9.1 & 0.0 & {\scriptsize   0692932001} \\
PSZ2 G324.04+48.79 & 13:47:30.5 & -11:45:07 & 0.4516 & 10.6 & 11.5 & 2 & 4.6 & 36.5 & 335.5 & 0.0 & {\scriptsize   0112960101} \\
PSZ2 G325.70+17.34 & 14:47:33.9 & -40:20:38 & 0.3155 & 7.6 & 8.7 & 2 & 6.8 & 0.0 & 30.1 & 45.0 & {\scriptsize   {\bf 0827020701}} \\
PSZ2 G339.63-69.34 & 23:44:43.9 & -42:43:11 & 0.5960 & 8.1 & 9.2 & 2 & 1.5 & 237.2 & 558.7 & 0.0 & {\scriptsize   0693661801 072270(01-02)01} \\
PSZ2 G340.36+60.58 & 14:01:02.2 & +02:52:43 & 0.2528 & 9.2 & 15.6 & 2 & 2.0 & 368.4 & 363.7 & 0.0 & {\scriptsize   0098010101 0147330201 055183(01-02)01} \\
PSZ2 G340.94+35.07 & 14:59:29.0 & -18:10:44 & 0.2357 & 7.8 & 10.5 & 2 & 7.4 & 0.0 & 40.2 & 24.0 & {\scriptsize   {\bf 0827311201}} \\
PSZ2 G346.61+35.06 & 15:15:03.1 & -15:22:46 & 0.2226 & 8.4 & 12.9 & 2 & 8.3 & 0.0 & 60.0 & 21.0 & {\scriptsize   {\bf 0827010201}$^{\star}$} \\
PSZ2 G349.46-59.95 & 22:48:44.4 & -44:31:58 & 0.3475 & 11.4 & 20.7 & 2 & 1.2 & 39.8 & 125.2 & 0.0 & {\scriptsize   0504630101} \\
\bottomrule
    % \hline
    \end{tabular}
    \caption{\footnotesize List of  \xmm\ observations.    \label{tab:master3b}}
\end{sidewaystable*}
%------------

%%%%%%%%%%%%%%%

\section{Weak lensing archive data}
\label{appx:wllist}

Tables~\ref{tab_lensing_data} and \ref{tab_lensing_data2} contain a summary of the archival optical data available for weak lensing (WL)  observations as of winter 2019. We only considered observations with an exposure time rescaled to an equivalent Subaru dish area longer than three minutes. We also list which clusters form part of WL samples in the literature: CLASH-WL are the CLASH clusters with measured WL mass from \citet{ume+al16b} or \citet{mer+al15}; WtG from \citet{wtg_III_14}; CCCP100 is the combined CCCP plus MENeaCS sample from \citet{her+al19}; LoCuSS from \citet{ok+sm16}; PSZ2LenS from \citet{ser+al17_psz2lens}; LC2 from $LC^2$  \citep{ser15_comalit_III}. The Table also contains information on WL surveys to be completed by 2022: the Hyper Suprime-Cam Subaru Strategic Program \citep[HSC-SSP]{hsc_aih+al18}; the CFHTLenS  \citep[Canada France Hawaii Telescope Lensing Survey,][]{hey+al12}; RCSLenS \citep[Red Cluster Sequence Lensing Survey,][]{hil+al16}; KiDS \citep[Kilo-Degree Survey ,][]{dej+al13}; DES \citep[ Dark Energy Survey ,][]{des16}. For ongoing surveys, we considered the final planned footprint.
The description of the relative broad band filters is presented in Table~\ref{tab_lensing_filters}.

%------------
% Lensing archival data
%
\begin{table*}[]
 \caption{\footnotesize Summary of archival data for weak lensing as of winter 2019. Columns 2-5: Available observations in multi-band filters at  worldwide facilities, see Table~\ref{tab_lensing_filters}. We only considered observations with an exposure time rescaled to an equivalent Subaru dish area longer than 3 minutes. Column 6: WL samples from literature; CLASH-WL are the CLASH clusters with measured WL mass from \citet{ume+al16b} or \citet{mer+al15}; WtG from \citet{wtg_III_14}; CCCP100 is the combined CCCP plus MENeaCS sample from \citet{her+al19}; LoCuSS from \citet{ok+sm16}; PSZ2LenS from \citet{ser+al17_psz2lens}; LC2 from $LC^2$  \citep{ser15_comalit_III}. Column 7: WL surveys to be completed by 2022: the Hyper Suprime-Cam Subaru Strategic Program \citep[HSC-SSP]{hsc_aih+al18}; the CFHTLenS  \citep[Canada France Hawaii Telescope Lensing Survey,][]{hey+al12}; RCSLenS \citep[Red Cluster Sequence Lensing Survey,][]{hil+al16}; KiDS \citep[Kilo-Degree Survey,][]{dej+al13}; DES \citep[ Dark Energy Survey,][]{des16}. For ongoing surveys, we considered the final planned footprint.}
    \label{tab_lensing_data}
    \centering
\resizebox{\hsize}{!} {
     \begin{tabular}{lllllllll} 
    %  \hline
    \toprule
    \toprule
     \noalign{\smallskip} 
 Name              	 & 	SuP@Subaru                 	 & 	HSC@Subaru     	 & 	 Megacam@CFHT 	 & 	OC@VST       	 & 	HAWKI    	 & 	WFI@MPG/ESO                                                              	 & 	WL samples                     	 & 	WL surveys       	\\
\midrule
PSZ2 G000.13+78.04  &  --  &  --  &  --  &  --  &  --  &  --  &  --  &  -- \\
PSZ2 G004.45-19.55  &  --  &  --  &  --  &  $i^+$  &  --  &  --  &  --  &  -- \\
PSZ2 G006.49+50.56  &  --  &  --  &  $g r i$  &  --  &  --  &  --  &  LC2,PSZ2LS,CCCP100  &  RCSLS \\
PSZ2 G008.31-64.74  &  --  &  --  &  --  &  $r^+ i^+$  &  $J$  &  --  &  --  &  KiDS \\
PSZ2 G008.94-81.22  &  $B_J R_C i^+ z^+$  &  --  &  $i$  &  $r^+ i^+$  &  $K_S$  &  $B_{99} B_{J,842} V_{843} V_{89} R_{C,844} R_{C,162} I_{C,879} z^{+,846}$  &  LC2  &  KiDS,DES \\
PSZ2 G021.10+33.24  &  $V_J i^+$  &  --  &  $g r$  &  --  &  --  &  $B_{J,842} R_{C,844} I_{C,879}$  &  LC2,LoCuSS,WtG,CCCP100  &  -- \\
PSZ2 G028.63+50.15  &  --  &  --  &  $r$  &  --  &  --  &  --  &  --  &  -- \\
PSZ2 G028.89+60.13  &  $V_J R_C i^+$  &  --  &  --  &  --  &  --  &  --  &  LC2,LoCuSS  &  -- \\
PSZ2 G031.93+78.71  &  --  &  --  &  $r$  &  --  &  --  &  --  &  --  &  -- \\
PSZ2 G033.81+77.18  &  --  &  --  &  $g r$  &  --  &  --  &  --  &  LC2,CCCP100  &  -- \\
PSZ2 G040.03+74.95  &  --  &  --  &  $r$  &  --  &  --  &  --  &  --  &  -- \\
PSZ2 G040.58+77.12  &  --  &  --  &  $r$  &  --  &  --  &  --  &  --  &  -- \\
PSZ2 G041.45+29.10  &  $V_J i^+$  &  --  &  --  &  --  &  --  &  --  &  --  &  -- \\
PSZ2 G042.81+56.61  &  --  &  $i^+$  &  $g r$  &  --  &  --  &  --  &  LC2,CCCP100  &  -- \\
PSZ2 G044.20+48.66  &  $g^+ R_C$  &  --  &  $g r$  &  --  &  --  &  --  &  LC2,CCCP100  &  -- \\
PSZ2 G044.77-51.30  &  $B_J V_J R_C I_C z^+$  &  --  &  $u r$  &  $r^+$  &  $J H$  &  --  &  LC2,WtG  &  -- \\
PSZ2 G046.10+27.18  &  $B_J V_J R_C I_C z^+$  &  --  &  $u r$  &  --  &  --  &  --  &  LC2,WtG  &  -- \\
PSZ2 G046.88+56.48  &  $V_J R_C i^+$  &  --  &  $g r$  &  --  &  --  &  --  &  LC2,CCCP100  &  -- \\
PSZ2 G048.10+57.16  &  $g^+ r^+ i^+$  &  --  &  $u g r$  &  --  &  --  &  --  &  --  &  CFIS \\
PSZ2 G049.22+30.87  &  $V_J R_C$  &  --  &  --  &  --  &  --  &  --  &  LC2,LoCuSS,WtG  &  -- \\
PSZ2 G049.32+44.37  &  --  &  --  &  $g r$  &  --  &  --  &  --  &  --  &  -- \\
PSZ2 G050.40+31.17  &  $V_J i^+$  &  --  &  $g r$  &  --  &  --  &  --  &  LC2,CCCP100  &  -- \\
PSZ2 G053.53+59.52  &  $B_J g^+ R_C z^+$  &  --  &  --  &  --  &  --  &  --  &  LC2  &  CFIS \\
PSZ2 G055.59+31.85  &  $V_J R_C i^+$  &  --  &  $u g r$  &  --  &  --  &  --  &  LC2,CLASH-WL,LoCuSS,WtG,CCCP100  &  CFIS \\
PSZ2 G056.77+36.32  &  --  &  --  &  --  &  --  &  --  &  --  &  LC2  &  CFIS \\
PSZ2 G056.93-55.08  &  $B_J V_J R_C I_C z^+$  &  --  &  $u g r i$  &  $r^+ i^+$  &  --  &  --  &  LC2,WtG  &  -- \\
PSZ2 G057.25-45.34  &  $B_J V_J R_C i^+ I_C z^+$  &  --  &  $u r$  &  $r^+$  &  --  &  --  &  LC2,WtG  &  -- \\
PSZ2 G057.61+34.93  &  --  &  --  &  $u g r$  &  --  &  --  &  --  &  --  &  CFIS \\
PSZ2 G057.78+52.32  &  --  &  --  &  --  &  --  &  --  &  --  &  --  &  CFIS \\
PSZ2 G057.92+27.64  &  --  &  --  &  --  &  --  &  --  &  --  &  --  &  CFIS \\
PSZ2 G062.46-21.35  &  --  &  --  &  $r$  &  --  &  --  &  --  &  --  &  -- \\
PSZ2 G066.41+27.03  &  $i^+$  &  --  &  --  &  --  &  --  &  --  &  --  &  CFIS \\
PSZ2 G066.68+68.44  &  $V_J i^+$  &  --  &  $i$  &  --  &  --  &  --  &  --  &  CFIS \\
PSZ2 G067.17+67.46  &  $g^+ V_J R_C i^+$  &  --  &  $g r$  &  --  &  --  &  --  &  LC2,LoCuSS,CCCP100  &  CFIS \\
PSZ2 G067.52+34.75  &  --  &  --  &  --  &  --  &  --  &  --  &  --  &  CFIS \\
PSZ2 G068.22+15.18  &  --  &  --  &  $r$  &  --  &  --  &  --  &  --  &  -- \\
PSZ2 G071.63+29.78  &  --  &  --  &  --  &  --  &  --  &  --  &  --  &  CFIS \\
PSZ2 G072.62+41.46  &  $B_J V_J R_C i^+$  &  --  &  --  &  --  &  --  &  --  &  LC2,LoCuSS,WtG,CCCP100  &  CFIS \\
PSZ2 G073.97-27.82  &  $B_J V_J R_C i^+ I_C z^+$  &  --  &  $u$  &  --  &  --  &  --  &  LC2,LoCuSS,WtG,CCCP100  &  -- \\
PSZ2 G075.71+13.51  &  $B_J V_J R_C i^+$  &  $i^+$  &  $g r$  &  --  &  --  &  --  &  --  &  -- \\
PSZ2 G077.90-26.63  &  $V_J i^+$  &  --  &  $g r$  &  --  &  --  &  --  &  --  &  -- \\
PSZ2 G080.16+57.65  &  --  &  --  &  --  &  --  &  --  &  --  &  --  &  CFIS \\
PSZ2 G080.37+14.64  &  --  &  --  &  $r$  &  --  &  --  &  --  &  --  &  -- \\
PSZ2 G080.41-33.24  &  --  &  --  &  $g r$  &  --  &  --  &  --  &  LC2,CCCP100  &  -- \\
PSZ2 G083.29-31.03  &  $B_J R_C I_C z^+$  &  --  &  $u g r i$  &  --  &  --  &  --  &  LC2,WtG  &  -- \\
PSZ2 G083.86+85.09  &  $V_J$  &  --  &  --  &  --  &  --  &  --  &  --  &  CFIS \\
PSZ2 G085.98+26.69  &  --  &  --  &  $r$  &  --  &  --  &  --  &  --  &  CFIS \\
PSZ2 G087.03-57.37  &  $B_J V_J R_C i^+$  &  --  &  $r i$  &  --  &  --  &  $B_{J,842} V_{843} R_{C,844}$  &  LC2,PSZ2LS,LoCuSS,WtG  &  HSC-SSP,RCSLS,DES \\
PSZ2 G092.71+73.46  &  $V_J i^+$  &  --  &  --  &  --  &  --  &  --  &  LC2,LoCuSS,CCCP100  &  CFIS \\
PSZ2 G094.69+26.36  &  --  &  --  &  --  &  --  &  --  &  --  &  --  &  CFIS \\
PSZ2 G098.44+56.59  &  --  &  --  &  $u g r i$  &  --  &  --  &  --  &  LC2,PSZ2LS  &  CFIS \\
PSZ2 G099.48+55.60  &  --  &  --  &  $u g r i$  &  --  &  --  &  --  &  LC2,PSZ2LS  &  CFIS,CFHTLS \\
PSZ2 G105.55+77.21  &  --  &  $i^+$  &  $u g i$  &  --  &  --  &  --  &  --  &  CFIS \\
PSZ2 G106.87-83.23  &  $V_J i^+$  &  --  &  --  &  --  &  --  &  $B_{J,842} B_{J,878} V_{843} R_{C,844}$  &  LC2,LoCuSS  &  DES \\
PSZ2 G107.10+65.32  &  $B_J g^+ V_J R_C i^+ z^+$  &  --  &  $g r$  &  --  &  --  &  --  &  LC2,LoCuSS,WtG,CCCP100  &  CFIS \\
PSZ2 G111.61-45.71  &  $B_J V_J R_C I_C$  &  $r^+ i^+ z^+ Y$  &  $u g r i$  &  --  &  --  &  --  &  LC2,WtG,CCCP100  &  -- \\
PSZ2 G111.75+70.37  &  --  &  --  &  --  &  --  &  --  &  --  &  --  &  CFIS \\
PSZ2 G113.29-29.69  &  --  &  --  &  $g r$  &  --  &  --  &  --  &  LC2,CCCP100  &  CFIS \\
PSZ2 G113.91-37.01  &  --  &  --  &  --  &  --  &  --  &  --  &  --  &  -- \\

    % \hline
    \bottomrule
    \end{tabular}
    }
 \end{table*}
 
 \begin{table*}[!h]
 \caption{\footnotesize Summary of archival data for weak lensing as of winter 2019 (continued).}
 \label{tab_lensing_data2}
     \centering
\resizebox{\hsize}{!} {
     \begin{tabular}{lllllllll} 
    %  \hline
    \toprule
    \toprule
     \noalign{\smallskip} 
   Name              	 & 	SuP@Subaru                 	 & 	HSC@Subaru     	 & 	 Megacam@CFHT 	 & 	OC@VST       	 & 	HAWKI    	 & 	WFI@MPG/ESO                                                              	 & 	WL samples                     	 & 	WL surveys       	\\
\midrule
PSZ2 G114.79-33.71  &  --  &  --  &  $g r$  &  --  &  --  &  --  &  LC2,CCCP100  &  -- \\
PSZ2 G124.20-36.48  &  $V_J i^+$  &  --  &  $g r$  &  --  &  --  &  --  &  LC2,LoCuSS,CCCP100  &  -- \\
PSZ2 G143.26+65.24  &  --  &  $i^+$  &  $r$  &  --  &  --  &  --  &  --  &  CFIS \\
PSZ2 G149.39-36.84  &  --  &  --  &  $r$  &  --  &  --  &  --  &  --  &  -- \\
PSZ2 G155.27-68.42  &  --  &  $r^+ i^+$  &  $u g r i$  &  --  &  --  &  --  &  --  &  DES \\
PSZ2 G159.91-73.50  &  $B_J V_J R_C i^+ z^+$  &  --  &  --  &  $r^+$  &  --  &  $B_{J,878} V_{843} I_{C,879}$  &  LC2,CLASH-WL,LoCuSS,WtG,CCCP100  &  DES \\
PSZ2 G172.74+65.30  &  --  &  --  &  --  &  --  &  --  &  --  &  --  &  CFIS \\
PSZ2 G172.98-53.55  &  $B_J V_J R_C i^+ I_C z^+ Y$  &  $z^+ Y$  &  $u g r i$  &  --  &  $J K_S$  &  $R_{C,844} I_{C,879}$  &  LC2,WtG,CCCP100  &  HSC-SSP,DES \\
PSZ2 G179.09+60.12  &  --  &  --  &  $g r$  &  --  &  --  &  --  &  LC2,CCCP100  &  CFIS \\
PSZ2 G186.37+37.26  &  $V_J i^+$  &  --  &  $g r$  &  --  &  --  &  --  &  LC2,LoCuSS,CCCP100  &  CFIS \\
PSZ2 G187.53+21.92  &  $V_J i^+$  &  --  &  $g r$  &  --  &  --  &  --  &  LC2,LoCuSS,CCCP100  &  -- \\
PSZ2 G192.18+56.12  &  --  &  --  &  $g r$  &  --  &  --  &  --  &  LC2,CCCP100  &  CFIS \\
PSZ2 G195.75-24.32  &  $V_J R_C i^+$  &  --  &  $g r$  &  --  &  --  &  $B_{99} V_{89} R_{C,844}$  &  LC2,CCCP100  &  -- \\
PSZ2 G201.50-27.31  &  $B_J V_J R_C i^+ I_C z^+$  &  --  &  $u g r i$  &  $r^+$  &  $J K_S H$  &  --  &  LC2,WtG,CCCP100  &  -- \\
PSZ2 G204.10+16.51  &  --  &  --  &  $u r$  &  --  &  --  &  --  &  --  &  -- \\
PSZ2 G205.93-39.46  &  $V_J R_C I_C$  &  --  &  $r$  &  $r^+$  &  $K_S$  &  --  &  LC2,WtG  &  -- \\
PSZ2 G206.45+13.89  &  --  &  --  &  $u r$  &  --  &  --  &  --  &  --  &  -- \\
PSZ2 G207.88+81.31  &  --  &  --  &  $r i$  &  --  &  --  &  --  &  --  &  -- \\
PSZ2 G208.80-30.67  &  $V_J R_C i^+ z^+$  &  --  &  $g r$  &  $r^+$  &  --  &  --  &  LC2,LoCuSS,WtG,CCCP100  &  -- \\
PSZ2 G210.64+17.09  &  --  &  --  &  $r$  &  --  &  --  &  --  &  --  &  -- \\
PSZ2 G216.62+47.00  &  $B_J V_J R_C i^+ I_C z^+$  &  $i^+$  &  $u$  &  --  &  --  &  --  &  LC2,WtG  &  -- \\
PSZ2 G217.09+40.15  &  --  &  $i^+$  &  $u g r i$  &  --  &  --  &  --  &  LC2,CCCP100  &  -- \\
PSZ2 G217.40+10.88  &  --  &  --  &  $r$  &  --  &  --  &  --  &  --  &  -- \\
PSZ2 G218.59+71.31  &  --  &  $i^+$  &  --  &  --  &  --  &  --  &  --  &  -- \\
PSZ2 G218.81+35.51  &  $V_J R_C i^+$  &  --  &  --  &  --  &  --  &  --  &  LC2,LoCuSS,WtG,CCCP100  &  -- \\
PSZ2 G224.00+69.33  &  $V_J i^+$  &  --  &  $g r$  &  --  &  --  &  --  &  LC2,CCCP100  &  -- \\
PSZ2 G225.93-19.99  &  --  &  $r^+ i^+$  &  --  &  --  &  $K_S$  &  --  &  --  &  -- \\
PSZ2 G226.18+76.79  &  $B_J V_J R_C i^+$  &  --  &  $u g r$  &  --  &  --  &  --  &  LC2,CCCP100  &  -- \\
PSZ2 G228.16+75.20  &  $B_J V_J R_C i^+ I_C z^+$  &  $z^+$  &  $u r$  &  --  &  $K_S$  &  --  &  LC2,CLASH-WL,WtG  &  -- \\
PSZ2 G229.74+77.96  &  --  &  $i^+$  &  --  &  --  &  --  &  --  &  --  &  -- \\
PSZ2 G238.69+63.26  &  $V_J i^+$  &  --  &  --  &  --  &  --  &  --  &  --  &  -- \\
PSZ2 G239.27-26.01  &  --  &  --  &  --  &  $r^+ i^+$  &  --  &  --  &  --  &  DES \\
PSZ2 G241.11-28.68  &  --  &  --  &  --  &  $r^+$  &  --  &  $R_{C,844}$  &  --  &  DES \\
PSZ2 G243.15-73.84  &  --  &  --  &  --  &  $r^+ i^+$  &  --  &  --  &  --  &  KiDS,DES \\
PSZ2 G243.64+67.74  &  --  &  --  &  $r$  &  --  &  --  &  --  &  --  &  -- \\
PSZ2 G259.98-63.43  &  --  &  --  &  --  &  $r^+ i^+$  &  --  &  $B_{99} B_{J,878} V_{843} V_{89} R_{C,844} R_{C,162} I_{C,879}$  &  LC2  &  DES \\
PSZ2 G262.27-35.38  &  --  &  --  &  --  &  $r^+$  &  --  &  $B_{J,842} B_{J,878} V_{843} R_{C,844} I_{C,879}$  &  LC2  &  DES \\
PSZ2 G262.73-40.92  &  --  &  --  &  --  &  --  &  --  &  --  &  --  &  DES \\
PSZ2 G263.68-22.55  &  --  &  --  &  --  &  $r^+$  &  --  &  $B_{J,841} B_{99} B_{J,878} V_{843} R_{C,844} I_{C,879}$  &  LC2  &  DES \\
PSZ2 G266.04-21.25  &  --  &  --  &  --  &  $r^+$  &  $J K_S$  &  $B_{J,842} B_{J,878} V_{843} R_{C,844} I_{C,845}$  &  LC2  &  DES \\
PSZ2 G266.83+25.08  &  --  &  $i^+$  &  --  &  $r^+$  &  --  &  --  &  --  &  -- \\
PSZ2 G271.18-30.95  &  --  &  --  &  --  &  $r^+$  &  $K_S$  &  --  &  --  &  DES \\
PSZ2 G273.59+63.27  &  --  &  --  &  --  &  $r^+ i^+$  &  --  &  $B_{J,842} V_{843} R_{C,844}$  &  LC2  &  HSC-SSP,KiDS \\
PSZ2 G277.76-51.74  &  --  &  --  &  --  &  --  &  --  &  $R_{C,844}$  &  --  &  DES \\
PSZ2 G278.58+39.16  &  $g^+ r^+$  &  --  &  $r$  &  $i^+$  &  --  &  $B_{J,842} B_{J,878} V_{843} R_{C,844}$  &  LC2  &  -- \\
PSZ2 G283.91+73.87  &  --  &  --  &  $u g r i$  &  --  &  --  &  --  &  --  &  -- \\
PSZ2 G284.41+52.45  &  $B_J V_J R_C I_C z^+$  &  --  &  $g r i$  &  $r^+$  &  $K_S$  &  $B_{J,878} V_{843} R_{C,844} I_{C,879}$  &  LC2,CLASH-WL,WtG  &  -- \\
PSZ2 G285.63+72.75  &  $V_J i^+$  &  --  &  $u g r i$  &  --  &  --  &  --  &  LC2  &  -- \\
PSZ2 G286.98+32.90  &  $g^+ r^+$  &  --  &  $g$  &  $g^+ i^+$  &  $K_S$  &  $V_{843} R_{C,844} I_{C,879}$  &  LC2  &  -- \\
PSZ2 G287.46+81.12  &  --  &  --  &  $r$  &  --  &  --  &  --  &  --  &  -- \\
PSZ2 G313.33+61.13  &  $B_J V_J R_C i^+ z^+$  &  --  &  --  &  $r^+ i^+$  &  $J K_S$  &  $B_{99} B_{J,878} V_{843} R_{C,844} R_{C,162} I_{C,879}$  &  LC2,LoCuSS  &  HSC-SSP,KiDS \\
PSZ2 G313.88-17.11  &  --  &  --  &  --  &  $r^+$  &  --  &  --  &  --  &  -- \\
PSZ2 G324.04+48.79  &  $V_J R_C I_C z^+$  &  --  &  $u g r i$  &  $r^+$  &  $K_S$  &  $B_{J,878} V_{843} R_{C,844} I_{C,879}$  &  LC2,CLASH-WL,WtG,CCCP100  &  -- \\
PSZ2 G325.70+17.34  &  --  &  --  &  --  &  --  &  --  &  --  &  --  &  -- \\
PSZ2 G339.63-69.34  &  --  &  --  &  --  &  --  &  --  &  --  &  --  &  DES \\
PSZ2 G340.36+60.58  &  $I_C$  &  --  &  $g r$  &  $r^+ i^+$  &  --  &  $B_{J,878} V_{843}$  &  LC2,LoCuSS,WtG,CCCP100  &  HSC-SSP,KiDS \\
PSZ2 G340.94+35.07  &  --  &  --  &  $i$  &  $r^+$  &  --  &  --  &  --  &  -- \\
PSZ2 G346.61+35.06  &  --  &  $i^+$  &  --  &  $g^+ r^+ i^+$  &  --  &  --  &  --  &  -- \\
PSZ2 G349.46-59.95  &  --  &  --  &  --  &  $g^+ r^+ i^+$  &  $K_S$  &  $B_{J,842} V_{843} R_{C,844} I_{C,879} z^{+,846}$  &  LC2,CLASH-WL  &  DES \\

    % \hline
    \bottomrule
    \end{tabular}
    }
 \end{table*}

\begin{table}[h!]
 \caption{Broad-band filter description}
 \label{tab_lensing_filters}
   \centering
%\resizebox{\hsize}{!} {
     \begin{tabular}{lll} 
\toprule
\toprule
    %  \hline 
	\multicolumn{1}{l}{Telescope/Instrument} & \multicolumn{1}{c}{Filter Name} & \multicolumn{1}{c}{Filter Description}  \\
% 	\hline
\midrule
 	\multicolumn{3}{l}{Subaru/Suprime-Cam} \\ 
 & $B_{\rm J}$ & Johnson $B$-band \\
 & $V_{\rm J}$ & Johnson $V$-band \\
 & $R_{\rm C}$ & Cousins $R$-band \\
 & $I_{\rm C}$ & Cousins $I$-band \\
 & $i^+$         &    SDSS $i$-band \\
 & $z^+$        &    SDSS $z$-band \\
\midrule
\multicolumn{3}{l}{ESO/WFI} \\ 
 & $U_{877}$ & $U$/50-band \\
 & $B_{842}$ & Johnson $B$-band \\
 & $V_{843}$ & Johnson $V$-band \\
 & $R_{844}$ & Cousins $R$-band \\
 & $I_{879}$ & Cousins $I$-band \\
 & $z_{846}$ &  $z{+}$/61-band \\
\midrule
\multicolumn{3}{l}{CFHT/MegaPrime} \\
& $u^\star$    &    SDSS $u$-band\\
& $g^\star$    &    SDSS $g$-band\\
& $r^\star$    &    SDSS $r$-band\\
& $i^\star$    &    SDSS $i$-band\\
& $z^\star$    &    SDSS $z$-band\\
\bottomrule
\end{tabular}
  %  }
\end{table}

%%%%%%%%%%%%%

\section{Available SZ data}
\label{appx:szlist}

By construction all clusters from our sample have {\it Planck} SZE data. They are available through the second public data release by the  {\it Planck} collaboration, which includes an all-sky SZ map \citep[i.e.  $y$-map --][]{PlanckDR15}.  {\it Planck} data are distributed though the  {\it Planck} Legacy Archive of ESA\footnote{\url{https://pla.esac.esa.int/pla/}}. SPT $y$-map data are not directly available yet for the full survey. However temperature maps (in units of $\mu$K-CMB ) of SPT data, either standalone or combined with  {\it Planck}, have been produced at each effective observing frequency of the instrument (i.e. 95, 150 and 220~GHz) and delivered as part the initial SPT release \citep[i.e. SPT-SZ --][]{cho18}. The ACT DR4 data have been delivered to the community. This release includes several types of $y$-maps over the ACT footprint (e.g. from the ACT data alone or in combination with  {\it Planck} or BOSS data -- \citealt{Aiola2020}).  ACT and SPT data are available through the Lambda\footnote{\url{https://lambda.gsfc.nasa.gov/}} portal from NASA. The Bolocam SZE data are publicly available as individual 140~GHz maps for each cluster \citep[in units of $\mu$K-CMB --][]{Sayers2013_pressure}\footnote{\url{https://irsa.ipac.caltech.edu/data/Planck/release_2/ancillary-data/bolocam/}}.  
The SZ NIKA data have been released by the NIKA Consortium \footnote{\url{http://lpsc.in2p3.fr/NIKA2LPSZ/nika2sz.release.php}}, whilst the NIKA2 data are still proprietary and shall be released by the NIKA2 consortium in the future.
All of the MUSTANG-1 cluster observations have been released in a similar manner\footnote{\url{https://safe.nrao.edu/wiki/bin/view/GB/Pennarray/MUSTANG_CLASH}}, although the MUSTANG-2 data are not yet publicly available. 
 The ALMA data are available through the Alma Science Archive portal\footnote{\url{https://almascience.eso.org/asax/}}.

%------------
% Table of SZE archival pointed data
%
\begin{table*}
\begin {center}
\caption{\footnotesize General characteristics of available SZE data and facilities.}
\label{tab:sz}
\resizebox{\hsize}{!} {
\begin{tabular}{lccccccr}
% \hline
% \hline
\toprule
\toprule
Facility or & Freq. Bands & Ang. Resolution$^a$ & Max Ang. Scale$^b$ & Coverage & Public DR$^c$ & Her. Gal. Clus. & Ref. \\
Survey           &        [GHz]   & [arcmin]   & [arcmin]     &  [sq. deg.]  & [year] & [number] \\   
% \hline
\midrule
Planck & 100,143,217,353,545,857 & 10.0$\phantom{0}$ & All Sky & All Sky &  $\phantom{>}$2015 & 118 & [1] \\
SPT-SZ & 95,150,220 & $\phantom{1}$1.75 & $\phantom{(+)}$11$(+)\phantom{.511}$ & $\phantom{1}$2,500 & $\phantom{>}$2018 & $\phantom{11}$9 & [2] \\
ACT-DR4 & 98, 150 & $\phantom{1}$1.4\phantom{1} & $\phantom{(+)}$11$(+)\phantom{.511}$ & 17,000 & $\phantom{>}$2020 & $\phantom{1}$56 & [3] \\
SPT-ECS & 95,150 & $\phantom{1}$1.2$\phantom{1}$ & $\phantom{(+)}$11$(+)\phantom{.511}$ & $\phantom{1}$2,770 & $\ge$2020 & $\phantom{1}$10 & [4]\\
Bolocam & 140 & $\phantom{1}$0.97 & $\phantom{(+)}$11$\phantom{.5(+)11}$ & -- & $\phantom{>}$2015 & $\phantom{1}$18 & [5]\\
NIKA-1 & 150,260 & $\phantom{1}$0.30 & $\phantom{(+)1}$4$\phantom{.5(+)11}$ & -- & $\phantom{>}$2017 & $\phantom{11}$1 & [6] \\
MUSTANG-1 & 90 & $\phantom{1}$0.15 & $\phantom{(+)1}$1.5$\phantom{(+)11}$ & -- & $\phantom{>}$2017 & $\phantom{11}$4 & [7]\\
NIKA2$^d$ & 150,260 & $\phantom{1}$0.30 & $\phantom{(+)1}$8$\phantom{.5(+)11}$ & -- & $\ge$2020 & $\phantom{11}$5 & [8,9]\\
MUSTANG-2$^d$ & 90 & $\phantom{1}$0.15 & $\phantom{(+)1}$4$\phantom{.5(+)11}$ & -- & -- & $\phantom{11}$1 & [10] \\
ALMA$^{d,e}$ & 92 & $\phantom{1}$0.07 & $\phantom{(+)1}$1.8$\phantom{(+)11}$ & -- & -- & $\phantom{11}$1 & [11]\\
% \hline \hline
\midrule
\multicolumn{5}{l}{\textbf{Total Number {\it XMM}-Heritage Galaxy Clusters with Ground-Based Coverage}} & Available Now & Future Release \\
\multicolumn{5}{l}{\hspace{0.25in} Tier-1} & 21$^f$ & $\phantom{1}$0 \\
\multicolumn{5}{l}{\hspace{0.25in} Tier-2} & 43$^f$ & $\phantom{1}$5 \\ 
% \hline
\bottomrule
\end{tabular}
}
\end {center}
    {\footnotesize{$^a$PSF FWHM, either for relevant released data from the surveys (e.g. the {\it Planck} $y$-map) or for the channel closest to 150~GHz; $^b$Maximum recoverable angular scale, corresponding to where the noise is approximately white for the ACT and SPT surveys (although the transfer function of each survey suggests that larger scales can likely be recovered) or the location where the transfer function equals 0.5 for the pointed instruments; $^c$Date of public release, or anticipated future release for SPT-ECS and the NIKA2 SZ Large Program; $^d$NIKA2, MUSTANG-2, and ALMA are available for open-time observations of additional {\it XMM}-Heritage galaxy clusters; $^e$The angular resolution and maximum angular scale for ALMA are based on the band~3 observations from \citet{kit16}; $^f$Includes one galaxy cluster that is in both Tier-1 and Tier-2. References: [1] \citep{planck2014-a01}, [2] \citet{cho18}, [3] \citet{Aiola2020}, [4] \citet{ble20}, [5] \citet{Sayers2013_pressure}, [6] \citet{Adam2014}, [7] \citet{Romero2017}, [8,9] \citet{rup18, Mayet2020}, [10] \citet{Romero2020}, [11] \citet{kit16}. }}

\end{table*}

\end{document}